\documentclass[twocolumn,superscriptaddress,aps,prd,nofootinbib,floatfix]{revtex4-2}
\usepackage{amssymb}
\usepackage{amsmath}
\usepackage{graphicx}
\usepackage{dcolumn}
\usepackage{bm}
\usepackage{dsfont}
\usepackage[breaklinks=true,colorlinks,citecolor=blue,linkcolor=blue,urlcolor=blue]{hyperref}
\usepackage{CJKutf8}
\usepackage{physics}
\usepackage{soul}
\usepackage{enumitem}
\usepackage{mleftright}
\usepackage{bbm}

\makeatletter
\renewcommand*\env@matrix[1][\arraystretch]{%
  \edef\arraystretch{#1}%
  \hskip -\arraycolsep
  \let\@ifnextchar\new@ifnextchar
  \array{*\c@MaxMatrixCols c}}
\makeatother

\newcommand{\be}{\begin{equation}}
\newcommand{\ee}{\end{equation}}
\newcommand{\bea}{\begin{eqnarray}}
\newcommand{\eea}{\end{eqnarray}}

\renewcommand{\eqref}[1]{\mbox{Eq.~(\ref{#1})}}

\newcommand{\figpanel}[2]{Fig.~\hyperref[#1]{\ref*{#1}(#2)}}

\usepackage{changes}
\definechangesauthor[name={Ale}, color=red]{Ale}
\definechangesauthor[name={David}, color=orange]{David}

\usepackage{tikz,xcolor}
\definecolor{lime}{HTML}{A6CE39}
\DeclareRobustCommand{\orcidicon}{%
	\begin{tikzpicture}
	\draw[lime, fill=lime] (0,0) 
	circle [radius=0.16] 
	node[white] {{\fontfamily{qag}\selectfont \tiny ID}};
	\draw[white, fill=white] (-0.0625,0.095) 
	circle [radius=0.007];
	\end{tikzpicture}
	\hspace{-2mm}
}
\foreach \x in {A, ..., Z}{%
	\expandafter\xdef\csname orcid\x\endcsname{\noexpand\href{https://orcid.org/\csname orcidauthor\x\endcsname}{\noexpand\orcidicon}}
}

\DeclareMathOperator{\sinc}{sinc}

\begin{document}
\title{Photon-graviton polarization entanglement induced by a classical electromagnetic wave}

\date{\today}

\author{Alessandro Ferreri\orcidA{}}
\affiliation{Institute for Quantum Computing Analytics (PGI-12), Forschungszentrum J\"ulich, 52425 J\"ulich, Germany}

\begin{abstract}
We study the photon-graviton pair production induced by the propagation of a classical electromagnetic (EM) wave in a Minkowskian spacetime. In our model, the gravitational field is described in terms of the quantized graviton field, whereas the electromagnetic field is split into a classical drive (a linearly or circularly polarized electromagnetic wave) and a quantum fluctuation field. We analyze the time evolution of the quantum state showing that, among other outcomes, the propagation of the EM wave can generate Bell states in the photon-graviton polarization basis. We finally discuss the possibility to observe entangled photons in artificial and natural scenarios.
\end{abstract}

\maketitle

\section{Introduction}
The fundamental nature of gravity is an enigma whose solution remains, at the present time, elusive \cite{dewittQuantumTheoryGravity1967,pfeiferModifiedQuantumGravity2023}. Despite all the extraordinary efforts of the last hundred years, we still possess neither a satisfactory quantum theory of gravity supported by experimental evidences, nor sufficient clues to claim with certainty the necessity of such theory at all \cite{oppenheimPostquantumTheoryClassical2023}. 
What we know hitherto is that we have a powerful classical apparatus, namely general relativity, that has correctly described the effects of the spacetime curvature, and has successfully predicted the propagation of gravitational signals. 

The existence of gravitational waves traveling across the universe is nowadays an ascertained fact. Since their first detection in 2016 by the LIGO project \cite{PhysRevLett.116.061102}, we have registered more than two hundred detection events \cite{Abbott_2017,Abbott_2020,PhysRevLett.125.101102,PhysRevD.102.043015, PhysRevX.13.011048}, allowing us to include gravitational waves among other cosmic information carriers like photons, cosmic rays and neutrinos; a discover which gave birth to the field of multi-messenger astronomy \cite{ADDAZI2022103948,piorkowska-kurpasGravitonMassEra2022}. 
The possibility to investigate cosmological phenomena via gravitational signals is an opportunity for us to better comprehend whether gravity only propagates as classical waves or, for example, it possesses quantum features, and it is therefore mediated by quanta called gravitons \cite{carneyGravitonDetectionQuantization2024, giovanniniMaximalFrequencyCosmic2024}.

In the last decade, different experimental proposals to test the quantum nature of gravity have been promoted and discussed \cite{dattaSignaturesQuantumNature2021,verlindeObservationalSignaturesQuantum2021,mehdiSignaturesQuantumGravity2023,PhysRevD.109.046012,PhysRevD.109.064078,carneyGravitonDetectionQuantization2024}. Two celebrated examples are \cite{boseSpinEntanglementWitness2017} and \cite{marlettoGravitationallyInducedEntanglement2017}, where the gravitational field non-locally mediates the interaction between two masses flowing in two spatially-separated interference set-ups. The claim of these proposals is therefore that the entanglement stemming from the nonlocality of the interaction is a signature of the quantum nature of the gravitational field, as only nonlocal operations can induce the entanglement between two subsystems \cite{RevModPhys.81.865,boseMechanismQuantumNatured2022}. 

Differently from the two proposals mentioned above, in which gravity mediates the entanglement occurring between two quantum objects (masses that possess quantum degrees of freedom \cite{boseMechanismQuantumNatured2022}), in this work we characterize the properties of a quantum system entangled with experimentally inaccessible quantum gravitational degrees of freedom. 
In order to fit gravity into a quantum framework and characterize the entanglement with the other quantum field, we restrict the focus on low energetic phenomena that slightly alter the spacetime structure, thereby expressing general relativity as a perturbative quantum field theory \cite{donoghueGeneralRelativityEffective1994, guptaQuantizationEinsteinGravitational1952, hsiangGravitonPhysicsConcise2024}. Although this approach inevitably fails when addressing ultrarelativistic processes or in proximity of extremely massive objects such as black holes, it offers valid insights in scenarios where the spacetime metric is weakly perturbed. 

In this work we therefore employ general relativity as a quantum field theory to analyse the interaction between the gravitational field and the electromagnetic field stimulated by the presence of a classical electromagnetic wave. We notice that similar scenarios have already been addressed in the fully classical, semiclassical, and quantum regime.
The study of classical aspects of such interactions relies on general relativity to examine, for example, the spacetime curvature originated by the propagation of a classical beam, or the mutual gravitational influence of two parallel beams \cite{tolmanGravitationalFieldProduced1931, scullyGeneralrelativisticTreatmentGravitational1979,ratzelGravitationalPropertiesLight2016,lageyreGravitationalInfluenceHigh2022}. On the other hand, semiclassical approaches, such as quantum field theory in curved spacetime, are based on the idea that the classical gravitational field is the background upon which the dynamics of quantum fields occurs \cite{birrell1984quantum}. Among other examples, semiclassical approaches are particularly helpful to study the propagation of photons in curved spacetime \cite{PhysRevD.90.045041,PhysRevD.104.085015}, or the degradation of entanglement in noninertial frames \cite{PhysRevLett.95.120404,PhysRevD.85.061701, PhysRevD.85.025012}.
At the quantum level, the interactions between the quantum electromagnetic field and the quantum gravitational field are usually intended as phenomena of photon-graviton scattering \cite{skobelevGravitonphotonInteraction1975,PhysRevD.74.124028,PhysRevD.98.023518}. Interestingly, the presence of strong, classical fields may alter the dynamics, facilitating for example photon-graviton conversion processes \cite{vjj8-xf7k}. It has been seen that these phenomena may occur in presence of a classical background characterized by sizable magnetic fields \cite{PhysRevLett.74.634,
PhysRevD.106.083508}.

Our analysis focuses on the production of photon-graviton pairs induced by a polarized electromagnetic beam traveling in a flat spacetime. We show that the polarization of the beam plays a crucial role on the dynamics, as it determines the polarization entanglement between photons and gravitons. Also, we characterize the state of the heralding photons, namely those photons witnessing the entanglement with quantum gravitational degrees of freedom. We conclude our study by discussing the possibility to detect heralding photons in controllable environment and in cosmological observations.

\section{Theoretical model}
\begin{figure}[h!]
    \centering
    \includegraphics[width=1\linewidth]{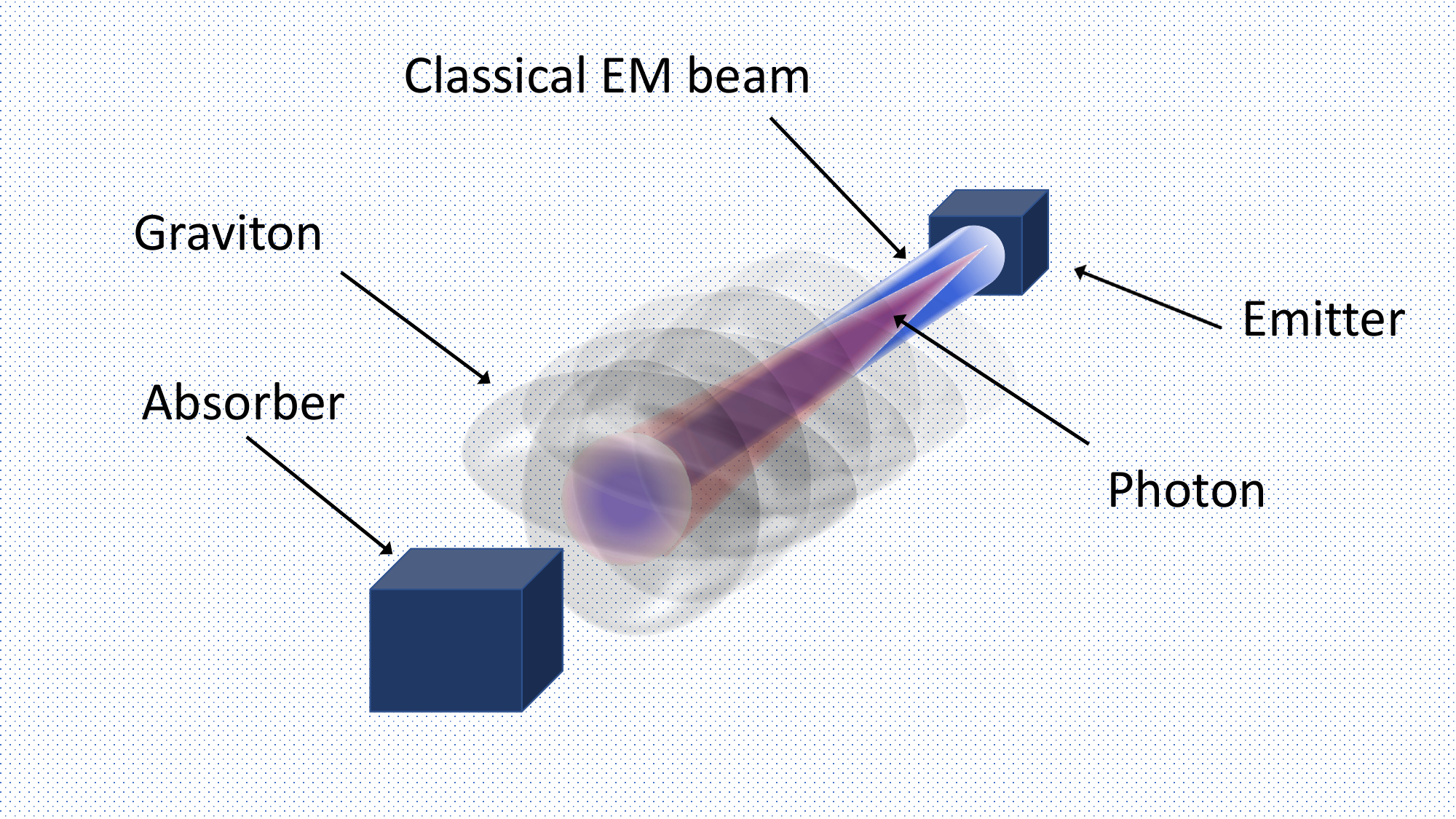}
    \caption{Pictorial representation of the scenario of interest: an intense classical electromagnetic beam propagates in the spacetime region between an emitter and an observer. Among other effects, we expect that the beam stimulates the vacua of both the quantum electromagnetic and the gravitational (graviton) fields, generating photon-graviton pairs.}
    \label{scheme}
\end{figure}

In this section, we present the theoretical model describing the interaction between the quantum electromagnetic field and the graviton field mediated by the presence of a classical EM wave. We recall that the main goal of our analysis is to extract information on the dynamics of the quantum state in order to investigate the presence of entanglement between quantum electromagnetic and gravitational degrees of freedom, see Fig.\ref{scheme}. Formally, we are therefore required to obtain a mathematical expression for the time evolution operator.

We start from the action describing both gravity and the electromagnetic field: 
\begin{align}
\mathcal{S}=\mathcal{S}_\textrm{EH}+\mathcal{S}_\textrm{em}
\end{align}
where we distinguish the Einstein-Hilbert action of the gravitational field,
\begin{align}
\mathcal{S}_\textrm{EH}=-\frac{c^2}{16\pi G}\int d^4x \sqrt{-g}R,
\end{align}
and the Maxwell action of the electromagnetic field,
\begin{align}
\mathcal{S}_\textrm{em}=-\frac{1}{4\mu_0}\int d^4x \sqrt{-g} F_{\mu\nu}F^{\mu\nu}.
\end{align}
In the expressions above, $g\equiv \textrm{det}(g_{\mu\nu})$ is the determinant of the spacetime metric $g_{\mu\nu}$, $R$ is the Ricci scalar, and 
\begin{equation}
F^{\mu\nu}=
\begin{pmatrix}[1.5]
0 & -\frac{ E_x}{c} & -\frac{ E_y}{c} & -\frac{ E_z}{c} \\
\frac{ E_x}{c} & 0 & - B_z&  B_y \\
\frac{ E_y}{c} &  B_z & 0& - B_x \\
\frac{ E_z}{c} & - B_y &  B_x& 0
\end{pmatrix},
\end{equation}
is the electromagnetic tensor, which encompasses the components of the electric $\bold E(\bold x)$ and magnetic field $\bold B(\bold x)$.

The mathematical procedure leading to the quantization of the gravitational field starts from the linearization of the spacetime metric \cite{flanaganBasicsGravitationalWave2005, capriniCosmologicalBackgroundsGravitational2018}. Far from gravitational sources, the metric can be written as the sum of a flat background and a small perturbation. In particular, we assume to work in a spacetime governed by the Minkowski metric perturbed by a small fluctuation: 
\begin{align}
g_{\mu\nu}=\eta_{\mu\nu}+h_{\mu\nu},
\end{align}
where $\eta=(-,+,+,+)$ is the standard Minkowski metric and $h$ is the perturbation. Note that henceforth tensor indices are raised and lowered via the Minkowski metric: $F_\mu{}^{\nu}\equiv\eta_{\mu\delta}F^{\delta\nu}$. Once expanded the Maxwell action, we can isolate the contribution to the Lagrangian density describing the interaction between the perturbation $h$ and the electromagnetic field. This reads
\begin{align}
\mathcal{L}_{(1)}=\frac{1}{2}h_{\mu\nu}T^{\mu\nu},
\label{lag}
\end{align}
where we introduced the stress energy tensor of the electromagnetic field
\begin{equation}
T^{\mu\nu}=
\begin{pmatrix}[1.5]
W & \frac{S_x}{c} & \frac{S_y}{c} & \frac{S_z}{c} \\
\frac{S_x}{c} & \sigma_{xx} & \sigma_{xy}& \sigma_{xz} \\
\frac{S_y}{c} & \sigma_{yx} & \sigma_{yy}& \sigma_{yz} \\
\frac{S_z}{c} & \sigma_{zx} & \sigma_{zy}& \sigma_{zz} 
\end{pmatrix},
\end{equation}
with $W=\frac{1}{2}\bold{E}^2+\frac{c^2}{2}\bold B^2$ the density energy,
$\bold S=c\bold{E}\times \bold B$
is the Poynting vector, and
$\sigma_{ij}=\delta_{ij}W-E_iE_j-c^2B_iB_j$ is the Maxwell stress tensor. The subscript (1) in \eqref{lag} indicates the first order expansion with respect to the perturbation $h$.

The electromagnetic field is stimulated by the presence of a EM beam traveling in the $z$ direction. Conveniently, we split the total electromagnetic field into a classical contribution due to the presence of the wave, and fluctuations whose quantum nature is described in terms of the quantized electromagnetic field.
The electric and magnetic fields are therefore written as
\begin{align}
\bold E(\bold x)&=\bm{\mathcal{E}}(\bold x)+\hat{\bold E}(\bold x)\nonumber\\
\bold B(\bold x)&=\bm{\mathcal{B}}(\bold x)+\hat{\bold B}(\bold x),
\label{classquant}
\end{align}
where $\bm{\mathcal{E}}$ and $\bm{\mathcal{B}}$ are the classical electric and magnetic field of the EM wave, respectively. Throughout this manuscript the EM wave will always be designed as a polarized, continuous-wave (CW), Gaussian beam traveling a distance $d$ in the $z$ direction. Since its explicit form depends on the polarization we choose, this will be given in a second moment.

The electric and magnetic field operators read
\begin{subequations}\label{quantfie}
\begin{align}
\hat{\bold E}(\bold x)=&i\sum_{\lambda=1}^2\int \frac{d^3\bold k}{\sqrt{2(2\pi)^3}} \sqrt{\hbar\omega_{\bold k}}\bold e^{\lambda\bold k}\left(\hat a_{\lambda\bold k}e^{i\bold k\cdot\bold x}+\textrm{h.c.}\right)\\
\hat{\bold B}(\bold x)=&i\sum_{\lambda=1}^2\int \frac{d^3\bold k}{\sqrt{2(2\pi)^3}} \bold k \times\bold e^{\lambda\bold k}\sqrt{\frac{\hbar}{\omega_{\bold k}}}\left(\hat a_{\lambda\bold k}e^{i\bold k\cdot\bold x}+\textrm{h.c.}\right)
\end{align}
\end{subequations}
where $\bold k$ is the wave vector, and the dispersion relation reads $\omega_{\bold k}=c\lvert \bold k\rvert$. The polarization vectors fulfill the orthogonal condition
$e^{i\lambda\bold k}e^{\bar\lambda\bold k}_i=\delta_{\lambda\bar\lambda}$
and the completeness relation
$\sum_{\lambda=1}^2   e^{\lambda\bold k}_ie^{\lambda\bold k}_j=P_{ij}^{\bold k}$
with $i,j=1,2,3$ and $P_{ij}^{\bold k}=\delta_{ij}-\frac{k_ik_j}{k^2}$. Throughout this manuscript we adopt the notation $\lambda=1\equiv\uparrow$ to indicate the vertical polarization and $\lambda=2\equiv\downarrow$ to indicate the horizontal polarization. The annihilation and creation operators fulfill standard bosonic commutation relations $[\hat a_{\lambda\bold k},\hat a_{\bar\lambda\bar{\bold k}}^\dag]=(2\pi)^3\delta(\bold k-\bar{\bold k})\delta_{\lambda\bar\lambda}$. The action of such operators on the vacuum state of the quantum electromagnetic field is $\hat a_{\lambda\bold k}\lvert 0_{\lambda\bold k}\rangle=0$ and $\hat a_{\lambda\bold k}^\dag\lvert 0_{\lambda\bold k}\rangle=\frac{1}{\sqrt{2\omega_{\lambda\bold k}}}\lvert 1_{\lambda\bold k}\rangle$, respectively \cite{Schwartz:2014sze}.

Once we cast the electromagnetic field expressed in \eqref{classquant} into the stress-energy tensor, we can split it into three components:
\begin{align}
\hat T^{\mu\nu}=T^{\mu\nu}_{\mathcal{E}_0^2}+\hat T^{\mu\nu}_{\mathcal{E}_0}+\hat T^{\mu\nu}_{\textrm{qf}}.
\label{Ttensor}
\end{align}
The first term only contains classical degrees of freedom, in particular it depends on the classical intensity of the EM wave $\mathcal{E}_0^2$. The second term is a mixture of classical parameters stemming from the EM wave and quantum operators, and it is linearly proportional to the classical amplitude $\mathcal{E}_0$. The third term does not contain any power of $\mathcal{E}_0$, and it is therefore negligible with respect to the first two terms. 

In this work we are not interested in studying the deformation of the spacetime caused by the presence of the EM wave. Backreaction effects onto the gravitational field due to the presence of classical electromagnetic waves have largely been studied in literature \cite{ratzelGravitationalPropertiesLight2016,lageyreGravitationalInfluenceHigh2022}. Here, we want to focus on the activation of particle production phenomena induced on a background of two quantum fields (graviton and photon field) by the presence of a classical drive. Since the EM wave is not responsible for the perturbation of the metric, the perturbation $h$ solves the Einstein equation in absence of source. We can therefore Fourier expand $h$ in terms of plane waves and express them in the transverse-traceless (TT) gauge, $h_{\mu}^\mu$=0, $h_{00}=h_{0i}\equiv0$ and $\partial_ih_{ij}=0$ \cite{flanaganBasicsGravitationalWave2005, capriniCosmologicalBackgroundsGravitational2018}.

The quantization of the gravitational degrees of freedom follows the standard protocol reported in literature \cite{Schwartz:2014sze}, and it leads to the quantized gravitational field \cite{guptaQuantizationEinsteinGravitational1952, hsiangGravitonPhysicsConcise2024}
\begin{align}
\hat h_{ij}(\bold r)=\mathcal{A}\sum_{\gamma=+\times}\int \frac{d^3\bold {p}}{\sqrt{2(2\pi)^3\omega_{\bold{p}}}}\epsilon_{ij}^{\gamma\bold p}\left(\hat b_{\gamma\bold p}e^{i\bold {p}\cdot\bold r}+\textrm{h.c.}\right)
\label{quangr}
\end{align}
where $\mathcal{A}=4\sqrt{\pi G\hbar}/c$ and $\bold p$ is the gravitational wave vector fulfilling the dispersion relation $\omega_{\bold{p}}=c\lvert\bold p\rvert$. 

Similarly to the electromagnetic field, the polarization vectors fulfill the orthogonal condition $\epsilon_{ij}^{\gamma\bold p}\epsilon^{\bar\gamma\bold pij}=\delta_{\gamma\bar\gamma}$ and the completeness relation $\sum_{\gamma}\epsilon_{ij}^{\gamma\bold p}\epsilon_{ls}^{\gamma\bold p}=\frac{1}{2}\left(P_{is}^{\bold p}P_{jl}^{\bold p}+P_{il}^{\bold p}P_{js}^{\bold p}-P_{ij}^{\bold p}P_{sl}^{\bold p}\right)$. The annihilation and creation operators for the graviton field fulfill the bosonic commutation relations $[\hat b_{\gamma\bold p},\hat b_{\bar\gamma\bar{\bold p}}^\dag]=(2\pi)^3\delta(\bold p-\bar{\bold p})\delta_{\gamma\bar\gamma}$. Their action on the gravitational vacuum state is $\hat b_{\gamma\bold p}\lvert 0_{\gamma\bold p}\rangle=0$ and $\hat b_{\gamma\bold p}^\dag\lvert 0_{\gamma\bold p}\rangle=\frac{1}{\sqrt{2\omega_{\bold p}}}\lvert 1_{\gamma\bold p}\rangle$, respectively.

The general expression of the Hamiltonian density describing the interactions between the electromagnetic and the graviton field have already been obtained, for example in \cite{mehdiSignaturesQuantumGravity2023}. Relevant steps of the calculations are reported in Appendix \ref{Hamandevo}.

\section{Photon-graviton entanglement}

The theoretical model discussed so far applies to any classical electromagnetic field. However, to proceed with the calculation of the interaction Hamiltonian, and consequently also of the time evolution operator, we need to characterize the classical electromagnetic field.

We recall that, in the interaction picture the time evolution of the state is performed by means of the time-evolution operator:
\begin{align}\label{time:evolution:operator}
\hat{U}_\textrm{I}(t)=&\overset{\leftarrow}{\mathcal{T}}\exp\left\{-\frac{i}{\hbar}\int_0^{t}dt'\hat H_\textrm{I}^{(\textrm{I})}(t')\right\},
\end{align}
where $\hat H_\textrm{I}^{(\textrm{I})}(t)$ is the interaction Hamiltonian in the interaction picture.
We write the unitary operator in the more convenient form $\hat{U}_\textrm{I}(t)=\hat{U}_{\mathcal{E}_0^2}(t)\hat{U}_{\mathcal{E}_0}(t)$, where
\begin{subequations}\label{Uop}
\begin{align}
\hat{U}_{\mathcal{E}_0^2}(t)=&\overset{\leftarrow}{\mathcal{T}}\exp\left\{-\frac{i}{\hbar}\int_0^{t}dt'\hat H_{\mathcal{E}_0^2}^{(\textrm{I})}(t')\right\},\label{UE2}\\
\hat{U}_{\mathcal{E}_0}(t)=&\overset{\leftarrow}{\mathcal{T}}\exp\left\{-\frac{i}{\hbar}\int_0^{t}dt'\hat{U}_{\mathcal{E}_0^2}^\dag(t')\hat H_{\mathcal{E}_0}^{(\textrm{I})}(t')\hat{U}_{\mathcal{E}_0^2}(t')\right\}\nonumber\\
\simeq& \hat{\mathbb{I}}-\frac{iC}{\hbar}\int_0^{t}dt'\hat{H}_{\textrm{eff}}(t')\nonumber\\
&-\frac{C^2}{\hbar^2}\int_0^t dt'\hat{H}_{\textrm{eff}}(t')\int_0^{t'} dt''\hat{H}_{\textrm{eff}}(t'').
\label{UE}
\end{align}    
\end{subequations}
In the expression above, we have perturbatively expanded $\hat{U}_{\mathcal{E}_0}(t)$ up to the second order with respect to a parameter $C$ which will be defined in a second moment. Moreover, we have introduced the effective Hamiltonian
\begin{align}
\hat{H}_{\textrm{eff}}(t)=\hat{U}_{\mathcal{E}_0^2}^\dag(t)\hat H_{\mathcal{E}_0}^{(\textrm{I})}(t)\hat{U}_{\mathcal{E}_0^2}(t).
\label{Heff}
\end{align} 

\subsection{Linearly polarized EM wave}
As a first scenario, we describe the EM wave as a classical electric field vertically polarized in the $x$ direction and a classical magnetic field oriented in the $y$ direction:
\begin{subequations}\label{classfie}
\begin{align}
\bm{\mathcal{E}}(\bold x,t)=&\bold e_x \mathcal{E}(\bold x,t)=\bold e_x\mathcal{E}_0 \xi(x,y)\sin(\Omega \,t-\kappa\,z)\\
\bm{\mathcal{B}}(\bold x,t)=&\bold e_y \mathcal{B}(\bold x,t)=\bold e_y \frac{\mathcal{E}_0}{c}\xi(x,y)\sin(\Omega \,t-\kappa\,z).
\end{align}
\end{subequations}
where $\mathcal{E}_0$ is the classical amplitude of the field, and the Gaussian function $\xi(x,y)=e^{-\frac{x^2+y^2}{4\sigma^2}}$ determines the spatial profile orthogonal to the direction of propagation $z$. As usual, the relation between frequency and wave vector is dictated by the dispersion relation $\Omega=c\kappa$, where $\bm{\kappa}\equiv(0,0,\kappa)$ is the wave vector of the EM wave.  The energy of the classical electromagnetic field is $\epsilon=\frac{1}{2}\int d^3\bold x \left(\left\lvert\mathcal{E}(\bold r,0)\right\rvert^2+\left\lvert\mathcal{B}(\bold r,0)\right\rvert^2\right)=\mathcal{E}_0^2V(1-\sinc(2\kappa d))$ where $V=\pi\sigma^2d$ is the effective volume of the beam.

We proceed with the calculation of the interaction Hamiltonian. In \eqref{Ttensor} we split the stress energy tensor into three terms proportional to different power of the classical wave amplitude $\mathcal{E}_0$. Substituting \eqref{quantfie} and \eqref{classfie} into \eqref{Ttensor} we have that the classical term of the stress energy tensor is 
\begin{equation}
\hat T^{\mu\nu}_{\mathcal{E}_0^2}=
\begin{pmatrix}
\frac{1}{2}\left(\mathcal{E}^2+c^2\mathcal{B}^2\right) & 0 & 0 & c\,\mathcal{E}\mathcal{B} \\
0 & 0 & 0& 0\\
0 & 0 & 0& 0 \\
c\,\mathcal{E}\mathcal{B} & 0 & 0& \frac{1}{2}\left(\mathcal{E}^2+c^2\mathcal{B}^2\right) 
\end{pmatrix},
\end{equation}
whereas the second term is
\begin{widetext}
\begin{equation}
\hat T^{\mu\nu}_{\mathcal{E}_0}=
\begin{pmatrix}
\mathcal{E}\hat E_x+c^2\mathcal{B}\hat B_y & -c\,\mathcal{B}\hat E_z & -c\,\mathcal{E}\hat B_z &c\left( \mathcal{E}\hat B_y+\mathcal{B}\hat E_x \right)\\
-c\,\mathcal{B}\hat E_z & -\mathcal{E}\hat E_x+c^2\mathcal{B}\hat B_y & -\mathcal{E}\hat E_y-c^2\mathcal{B}\hat B_x & -\mathcal{E}\hat E_z\\
-c\,\mathcal{E}\hat B_z & -\mathcal{E}\hat E_y-c^2\mathcal{B}\hat B_x & \mathcal{E}\hat E_x-c^2\mathcal{B}\hat B_y& -c^2\mathcal{B}\hat B_z \\c\left(
\mathcal{E}\hat B_y+\mathcal{B}\hat E_x\right) & -\mathcal{E}\hat E_z & -c^2\mathcal{B}\hat B_z& \mathcal{E}\hat E_x+c^2\mathcal{B}\hat B_y 
\end{pmatrix}.
\end{equation}
\end{widetext}
In the expressions above we have conveniently simplified the notation: $\mathcal{E}\equiv\mathcal{E}(\bold x,t)$ and $\mathcal{B}\equiv\mathcal{B}(\bold x,t)$. 
We also recall that we suppressed the third term in \eqref{Ttensor}.
By means of \eqref{pi1} and \eqref{HI}, and noting that $\bm{\mathcal{E}}^2=c^2\bm{\mathcal{B}}^2$, the two terms of
\eqref{hamden} for a EM wave polarized along the $x$ direction become

\begin{subequations}\label{hdens}
\begin{align}
\mathcal{H}_{\mathcal{E}_0^2}=&\frac{1}{2}\left(\hat h_{22}-\hat h_{11}\right)\mathcal{E}^2,\\
\mathcal{H}_{\mathcal{E}_0}=&\left[\left(\hat h_{22}+\hat h_{33}\right)\hat E_x-\hat h_{12}\hat E_y-\hat h_{13}\hat E_z\right]\mathcal{E}\nonumber\\
&-c\left[\left(\hat h_{11}+\hat h_{33}\right)\hat B_y-\hat h_{12}\hat B_x-\hat h_{23}\hat B_z\right]\mathcal{E}.
\end{align}
\end{subequations}
Note that the Hamiltonian density $\mathcal{H}_{\mathcal{E}_0^2}$ only depends on quantum operators of the gravitational field. Therefore, the corresponding Hamiltonian $\hat H_{\mathcal{E}_0^2}$ is a quantum operator that acts only on the gravitational quantum degrees of freedom. 
In the next section, we calculate the explicit forms of $\hat H_{\mathcal{E}_0^2}$ and $\hat H_{\mathcal{E}_0}$ and the time evolution operators $\hat U_{\mathcal{E}_0^2}$ and $\hat U_{\mathcal{E}_0}$ defined in \eqref{Uop}.

\subsubsection{Calculations of $\hat H_{\mathcal{E}_0^2}$ and $\hat H_{\mathcal{E}_0}$}

The first step consists in replacing the graviton and the electromagnetic fields in \eqref{hdens} with their explicit expressions given in \eqref{quangr}, \eqref{quantfie} and \eqref{classfie}, respectively.
To obtain the interaction Hamiltonians $\hat H_{\mathcal{E}_0^2}$ and $\hat H_{\mathcal{E}_0}$ from \eqref{hdens}, we are then required to compute spatial integrals. 

We start from $\hat H_{\mathcal{E}_0^2}$. The integrals in the $x$ and $y$ directions give
\begin{align}
\int dxdy \,e^{  i(p_x x+p_y y)}\xi^2(x,y)=2\pi\sigma^2\,e^{-\left(p_x^2+p_y^2\right)\frac{\sigma^2}{2}}
\end{align}
and along $z$ we have
\begin{align}
f(p_z,t)\equiv&\int_0^ddz e^{ip_z z}\sin^2\left(\kappa z-\Omega t\right)\nonumber\\
=&\frac{i\,e^{i p_z d}}{2p_z}\left[\frac{p_z^2\cos(2q)}{p_z^2-4\kappa^2}-\frac{2i\,\kappa p_z\sin(2q)}{p_z^2-4\kappa^2}-1\right]\nonumber\\
&+\frac{i}{2p_z}\left[1-\frac{p_z^2\cos(2\Omega t)}{p_z^2-4\kappa^2}-\frac{2i\,\kappa p_z\sin(2\Omega t)}{p_z^2-4\kappa^2}\right]
\end{align}  
where $q=\kappa d-\Omega t$.
Carried out the spatial integrals we find that the Hamiltonian $\hat H_{\mathcal{E}_0^2}$ in the interaction picture is
\begin{align}
\hat H_{\mathcal{E}_0^2}^{(\textrm{I})}(t)=&\frac{\mathcal{E}_0^2 \mathcal{A}\sigma^2}{4\sqrt{\pi}}\sum_\gamma\int \frac{d^3\bold {p}}{\sqrt{\omega_{\bold{p}}}}e^{-\left(p_x^2+p_y^2\right)\frac{\sigma^2}{2}}\left(\epsilon_{yy}^{\gamma\bold p}-\epsilon_{xx}^{\gamma\bold p}\right)\nonumber\\
&\left[f(p_z,t)e^{-i\,\omega_{\bold {p}} t}\hat b_{\gamma\bold p}+\textrm{h.c.}\right].
\label{hE2int}
\end{align}

The presence of the classical wave modulates the Hamiltonian, which acts as an external drive on the gravitational degrees of freedom. However, it is relevant to notice that not all gravitational modes are effectively driven by $\hat H_{\mathcal{E}_0^2}$. Indeed, the modulation caused by oscillating terms in $f(p_z,t)$ selects those modes whose frequency spans in a range closed to $2\Omega$. By means of the rotating wave approximation \cite{Heib_2025}, we can therefore discard all secularities, namely the counterrotating terms, and reduce the Hamiltonian in \eqref{hE2int} to
\begin{align}
\hat H_{\mathcal{E}_0^2}^{(\textrm{I})}(t)=&\frac{\mathcal{E}_0^2 \mathcal{A}\sigma^2d}{16\sqrt{\pi}}\sum_\gamma\int \frac{d^3\bold {p}}{\sqrt{\omega_{\bold{p}}}}\left(\epsilon_{xx}^{\gamma\bold p}-\epsilon_{yy}^{\gamma\bold p}\right)e^{-i(\omega_{\bold{p}}-2\Omega)t}\nonumber\\
&e^{-\left(p_x^2+p_y^2\right)\frac{\sigma^2}{2}}\sinc\left[(p_z-2\kappa)\frac{d}{2}\right]e^{i(p_z-2\kappa)\frac{d}{2}}\hat b_{\gamma\bold p}\nonumber\\
&+\textrm{h.c.}
\label{HE2}
\end{align}
In the limits $\kappa\sigma\gg1$ and $\kappa d\gg1$, we can replace the Gaussian and sinc functions to Dirac delta functions: $\sigma e^{-\frac{X^2\sigma^2}{2}}\rightarrow \sqrt{2\pi}\delta(X)$ and $d\sinc\left(\frac{X d}{2}\right)\rightarrow 2\pi\delta(X)$. Therefore, we obtain 
\begin{align}
\hat H_{\mathcal{E}_0^2}^{(\textrm{I})}(t)=&\frac{\mathcal{E}_0^2 \mathcal{A}\pi^2}{4\sqrt{\pi\Omega}}\int d^3\bold {p}\,\delta(\bold{p}-2\kappa)\left(e^{i(2\Omega-\omega_{\bold{p}})t}\hat b_{+\bold p}+\textrm{h.c.}\right)\nonumber\\
=&\frac{\mathcal{E}_0^2 \mathcal{A}\pi^2}{4\sqrt{\pi\Omega}}\left(\hat b_{+2\bm\kappa}+\hat b_{+2\bm\kappa}^\dag\right),
\label{HE2lim}
\end{align}
where we adopted the notation $
\hat b_{+2\bm\kappa}\equiv\int d^3\bold {p}\,\delta(\bold{p}-2\bm\kappa)\hat b_{+\bold p}$.

We observe that the delta function selects the mode whose wave vector $\bold p$ is $(0,0,2\kappa)$. The stimulated graviton therefore propagate in the $z$ direction with frequency $\omega_\bold{p}=2\Omega$. 
The polarization $+$ of the stimulated mode is a consequence of the factor $\left(\epsilon_{xx}^{\gamma\bold p}-\epsilon_{yy}^{\gamma\bold p}\right)$ in \eqref{HE2}. In the TT gauge, the polarization tensors of a graviton propagating along the $z$ direction are \cite{hsiangGravitonPhysicsConcise2024}
\begin{equation}
\epsilon_{\mu\nu}^{+\bold p}=
\begin{pmatrix}
0 & 0 & 0 & 0 \\
0 & 1 & 0& 0\\
0 & 0 & -1& 0 \\
0 & 0 & 0& 0 
\end{pmatrix}, \;\;\textrm{and}\;\; 
\epsilon_{\mu\nu}^{\times\bold p}=\begin{pmatrix}
0 & 0 & 0 & 0 \\
0 & 0 & 1& 0\\
0 & 1 & 0& 0 \\
0 & 0 & 0& 0 
\end{pmatrix}.
\end{equation}
Since only the polarization tensor $\epsilon_{\mu\nu}^{+\bold p}$ possesses the components $xx$ and $yy$, we can safely claim that the generated graviton is $+$polarized, and the factor $\left(\epsilon_{xx}^{\gamma\bold p}-\epsilon_{yy}^{\gamma\bold p}\right)$ reduces to 2.

Note that the Hamiltonian in \eqref{HE2lim} becomes time-independent as a result of the rotating wave approximation. The unitary operator in \eqref{UE2} is therefore easy to calculate:
\begin{align}
\hat{U}_{\mathcal{E}_0^2}(t)\equiv&\overset{\leftarrow}{\mathcal{T}}\exp\left\{-\frac{i}{\hbar}\int_0^{t}dt'\hat H_{\mathcal{E}_0^2}^{(\textrm{I})}(t')\right\}=e^{-\frac{i}{\hbar}\mathcal{H}_{\mathcal{E}_0^2}^{(\textrm{I})}t}\nonumber\\
=&\hat D_{+2\kappa}[\alpha(t)],
\label{displ}
\end{align}
where we defined the displacement operator $\hat D_X[\alpha(t)]=\exp\{\alpha(t) \hat b_X^\dag-\alpha^*(t)\hat b_X\}$,
with time-dependent displacement parameter $\alpha(t)=\alpha_0 t=-\frac{it\mathcal{E}_0^2 \mathcal{A}\pi^2}{4\hbar\sqrt{\pi\Omega}}$. The action of the displacement operator on the graviton annihilation operator is
\begin{align}
\hat D_{+2\kappa}^\dag[\alpha] \hat b_{\lambda\bold p}\hat D_{+2\kappa}[\alpha]=\hat b_{\lambda\bold p}+(2\pi)^3\alpha\,\delta(\bold p-2\kappa)\delta_{\lambda+}.
\end{align}

We highlight the fact that a classical counterpart of the coherent stimulation of the gravitational field, in which the propagation of a classical EM wave generates an oscillatory perturbation of the spacetime with twice the frequency of the EM wave, has already been predicted in \cite{lageyreGravitationalInfluenceHigh2022}. In the fully classical scenario, the solution to the Einstein equation for a spacetime perturbatively altered by the presence of a EM wave contains a constant part and an oscillating contribution \cite{lageyreGravitationalInfluenceHigh2022}. Therefore, the displacement of the graviton mode here obtained reproduces the reaction of the gravitational field to the stimulation of the propagating wave at the quantum scale. We also notice that our gauge choice allows us to only access the information on the oscillating solution of the Einstein equation, as this corresponds to the propagating gravitational signal. Therefore, our results cannot reproduce any quantum equivalents to the constant part of the gravitational field. 

Finally, we stress that the quantum interpretation of the result obtained in \cite{lageyreGravitationalInfluenceHigh2022} here proposed, namely the generation of coherent gravitons caused by the propagation of an EM wave, can hardly be exploited as a test for the quantum features of the gravitational field, as coherent states display classical features in the limit $\alpha\rightarrow\infty$ \cite{zhangCoherentStatesTheory1990}. On the other hand, the quantum model here discussed also predicts graviton-photon entanglement phenomena that can only arise assuming quantum degrees of freedom for the gravitational field.

Now we focus on $\hat H_{\mathcal{E}_0}$.
Once computed the integrals along the $x$ and $y$ direction we have
\begin{align}
&\int dx dy \xi(x,y)e^{ i(k_x+ p_x)x}e^{i(k_y+ p_y)y}\nonumber\\
=&4\pi\sigma^2 e^{-(k_x+ p_x)^2\sigma^2}e^{-(k_y+ p_y)^2\sigma^2}
\end{align}
and along $z$ we obtain
\begin{align}
 g(p_z+k_z,t)\equiv&\int_0^d dz e^{i(k_z+p_z)z}\sin\left(\Omega t-\kappa z\right)\nonumber\\
=&\frac{\kappa \cos(\Omega t)+i(k_z+p_z)\sin(\Omega t)}{(k_z+p_z)^2-{\kappa}^2}\nonumber\\
&-\frac{i(k_z+p_z)e^{i(k_z+p_z)d}\sin(\Omega t-\kappa d)}{(k_z+p_z)^2-{\kappa}^2}\nonumber\\
&-\frac{\kappa\,e^{i(k_z+p_z)d} \cos(\Omega t-\kappa d)}{(k_z+p_z)^2-{\kappa}^2},
\end{align}
therefore the Hamiltonian is
\begin{align}
\hat H_{\mathcal{E}_0}=&\frac{i\sqrt\hbar\mathcal{E}_0\mathcal{A}\sigma^2}{(2\pi)^2}\nonumber\\
&\sum_{\lambda,\gamma}\int d^3\bold p\,d^3\bold k\sqrt{\frac{\omega_{\bold k}}{\omega_{\bold p}}}\, \mathcal{I}_{\lambda\bold k}^{\gamma\bold p}e^{-\left[(k_x+ p_x)^2+(k_y+ p_y)^2\right]\sigma^2}\nonumber\\
&\left[\hat a_{\lambda\bold k} g(p_z+k_z,t)-\hat a_{\lambda\bold k}^\dag g(p_z-k_z,t)\right]\hat b_{\gamma\bold p}+\textrm{h.c.}
\end{align}
where
\begin{align}
\mathcal{I}_{\lambda\bold k}^{\gamma\bold p}=&
\left(\epsilon^{\gamma\bold p}_{yy}+\epsilon^{\gamma\bold p}_{zz} \right)e^{\lambda\bold k}_x-\epsilon^{\gamma\bold p}_{xy} e^{\lambda\bold k}_y- \epsilon^{\gamma\bold p}_{xz} e^{\lambda\bold k}_z\nonumber\\
&+\epsilon^{\gamma\bold p}_{yz} (\bar k_xe^{\lambda\bold k}_y-\bar k_ye^{\lambda\bold k}_x)+ \epsilon^{\gamma\bold p}_{xy} (\bar k_ye^{\lambda\bold k}_z-\bar k_ze^{\lambda\bold k}_y)\nonumber\\
&-\left(\epsilon^{\gamma\bold p}_{xx} +\epsilon^{\gamma\bold p}_{zz} \right)(\bar k_z e^{\lambda\bold k}_x-\bar k_x e^{\lambda\bold k}_z)
\label{Ioriginal}
\end{align}
is a function that depends on the components of the polarization tensor of the gravitational field, as well as on the components of both the wave vector and the polarization vector of the quantum electromagentic field.
The bar indicates the normalization with respect to the modulus of the wave vector: $\bar k_i\equiv k_i/\lvert k\rvert$.

It is relevant to notice that similar Hamiltonian operators have largely been studied in other areas of physics. In quantum optics and optomechanics, such Hamiltonian operators can describe the interactions between two bosonic modes controlled by an external drive. Phenomena of this kind are, for example, laser cooling processes via modulated Raman scattering \cite{RevModPhys.86.1391}, and the parametric down-conversion (PDC) \cite{PhysRevA.31.2409,couteau2018spontaneous}. The latter describes the single mode or two-mode squeezing effect occurring in a nonlinear media stimulated by the propagation of a laser beam. Interestingly, photons generated via the PDC effect can be subject to different form of entanglement \cite{PhysRevA.79.053842}, such as spectral \cite{PhysRevA.62.043816}, spatial \cite{PhysRevLett.64.2495} and polarization \cite{PhysRevLett.75.4337} entanglements. In an analog manner, we will not only see that the propagation of the classical EM wave can induce the pair production of one photon and one graviton, but also that the generated pair is subject to polarization entanglement.

Once again, we move our mathematical description to the interaction picture and perform the rotating wave approximation. This allows us to exclude counterrotating terms containing the oscillating factors $e^{\pm i\left(\Omega+\omega_{\bold k}+\omega_{\bold p}\right)t}$. We are therefore left with
\begin{align}
\hat H_{\mathcal{E}_0}^{(\textrm{I})}(t)=&\frac{\sqrt\hbar\mathcal{E}_0\mathcal{A}V}{(2\pi)^3}\sum_{\lambda,\gamma}\int d^3\bold p\,d^3\bold k\sqrt{\frac{\omega_{\bold k}}{\omega_{\bold p}}}\,\mathcal{I}_{\lambda\bold k}^{\gamma\bold p}\,e^{i\left(\Omega-\omega_{\bold p}\right)t}\nonumber\\
& \left\{\hat a_{\lambda\bold k}\chi(\bold k_\perp+\bold p_\perp,k_z+p_z)e^{-i\omega_{\bold k}t}\right.\nonumber\\
&\left.-\hat a_{\lambda\bold k}^\dag \chi(\bold k_\perp+\bold p_\perp,-k_z+p_z)e^{i\omega_{\bold k}t}\right\}\hat b_{\gamma\bold p}+\textrm{h.c.}
\label{He0}
\end{align}   
where
\begin{align}
\chi(\bold k_\perp,k_z)=&e^{-(k_x^2+k_y^2)\sigma^2}e^{i\frac{(\kappa-k_z) d}{2}}\sinc\left[\left(k_z-\kappa\right)\frac{d}{2}\right].
\label{chi}
\end{align}
The notation $\bold k_\perp$ stands for the vector $(k_x,k_y)$. Such Hamiltonian finally reduces to
\begin{widetext}
\begin{align}
\hat H_{\mathcal{E}_0}^{(\textrm{I})}(t)=&C\sum_{\lambda,\gamma}\int d^3\bold p\,d^3\bold k\sqrt{\frac{\omega_{\bold k}}{\omega_{\bold p}}}\,\mathcal{I}_{\lambda\bold k}^{\gamma\bold p} \delta(k_x+p_x)\delta(k_y+p_y)e^{-i\left(\kappa-p_z \right)\frac{d}{2}}e^{i\left(\Omega-\omega_{\bold p}\right)t}\nonumber\\
& \left\{\hat a_{\lambda\bold k}e^{\frac{ik_zd}{2}}e^{-i\omega_{\bold k}t}\delta(\kappa-k_z-p_z)-\hat a_{\lambda\bold k}^\dag e^{-\frac{ik_zd}{2}}e^{i\omega_{\bold k}t}\delta(\kappa+k_z-p_z)\right\}\hat b_{\gamma\bold p}+\textrm{h.c.}
\label{He0-2}
\end{align}
\end{widetext}
in the limits $\kappa\sigma\gg1$ and $\kappa d\gg1$ we have $\sigma e^{-X^2\sigma^2}\rightarrow \sqrt{\pi}\delta(X)$ and $d\sinc\left(\frac{X d}{2}\right)\rightarrow 2\pi\delta(X)$,
with $C=\frac{\sqrt\hbar\mathcal{E}_0\mathcal{A}}{4}$.

The Hamiltonian in \eqref{He0} is required to calculate the effective Hamiltonian in \eqref{Heff} and the time evolution operator $\hat{U}_{\mathcal{E}_0}(t)$ in \eqref{UE}. Once applied the displacement operator \eqref{displ} to \eqref{He0} we obtain 
\begin{align}
\hat{H}_{\textrm{eff}}(t)=\hat H_{\mathcal{E}_0}^{(\textrm{I})}(t)+\hat H_{\kappa}(t),
\label{Heff2}
\end{align}
where we have defined the Hamiltonian
\begin{align}
\hat H_{\kappa}(t)=&\frac{\sqrt\hbar\mathcal{E}_0\mathcal{A} V\alpha(t)}{\sqrt{2\Omega}}\,\sum_{\lambda=1}^{2}
\int d^3\bold k\,\sqrt{\omega_{\bold k}}\,\mathcal{I}_{\lambda\bold k}^{+2\bold\kappa}e^{-i\left(\Omega+\omega_{\bold k}\right)t}\nonumber\\
&\left[\hat a_{\lambda\bold k}\chi(\bold k_\perp,-k_z)-\hat a_{\lambda\bold k}^\dag e^{2i\omega_{\bold k}t}\chi(\bold k_\perp,k_z)\right]+\textrm{h.c.}
\end{align}
The rotating wave approximation simplifies the expressions above, as it gets rid of $\chi(\bold k_\perp,-k_z)$. This leads to the Hamiltonian
\begin{align}
\hat H_{\kappa}(t)=&-\frac{4C\,V \alpha(t)}{\sqrt{2\Omega}}\nonumber\\
&\sum_{\lambda=1}^{2}
\int d^3\bold k\,\sqrt{\omega_{\bold k}}\mathcal{I}_{\lambda\bold k}^{+2\bold\kappa}e^{-i\left(\Omega-\omega_{\bold k}\right)t}\hat a_{\lambda\bold k}^\dag\chi(\bold k_\perp,k_z)\nonumber\\
&+\textrm{h.c.}
\label{Hkappa}
\end{align}
which reduces to
\begin{align}
\hat H_{\kappa}(t)=&-4\pi^3\sqrt{2}C\left[\hat a_{\uparrow\bm\kappa}^\dag \alpha(t)+\hat a_{\uparrow\bm\kappa}\alpha^*(t)\right]
\label{Hkappa2}
\end{align}
in the usual limits $\kappa\sigma\gg1$ and $\kappa d\gg1$. As before, we adopted the notation $
\hat a_{\uparrow\bm\kappa}\equiv\int d^3\bold {k}\,\delta(\bold{k}-\bm\kappa)\hat a_{\uparrow\bold k}$.

We notice that the Hamiltonian $\hat H_{\kappa}(t)$ has a clear physical interpretation. It describes the backreaction of the quantum electromagnetic field in response to the graviton displacement induced by the EM wave. In other words, the EM wave generates coherent gravitons, which in turn decay into photons whose frequency and polarization match those of the EM wave. 

To obtain the explicit form of the time evolution operator we need to perform the time integral of $\hat H_{\kappa}(t)$ and $\hat H_{\mathcal{E}_0}^{(\textrm{I})}(t)$. This leads to
\begin{align}
\int_0^t dt' \hat H_{\kappa}(t')=&\frac{t}{2}\hat H_{\kappa}(t),
\label{Hkappaint}
\end{align}
and
\begin{widetext}
\begin{align}
\int_0^t dt' \hat H_{\mathcal{E}_0}^{(\textrm{I})}(t')=&\frac{(2\pi)^2\sqrt{2} C}{\Omega\,c^3}\left\{\int_0^\infty d\omega \,\sqrt{\omega^3}\sqrt{\omega+\Omega}\left[\hat a_{\uparrow\omega}^\dag \hat b_{+(\Omega+\omega)}+\hat a_{\downarrow\omega}^\dag \hat b_{\times(\Omega+\omega)}\right]\right.\nonumber\\
&\left.-\int_0^\Omega d\omega\,\sqrt{\omega^3}\sqrt{\Omega-\omega}\left[\hat a_{\uparrow\omega}\hat b_{+(\Omega-\omega)}+\hat a_{\downarrow\omega}\hat b_{\times(\Omega-\omega)}\right]\right\}+\textrm{h.c.}
\label{intHe0}
\end{align}
\end{widetext}
More details on the calculations of such integrals are found in Appendix \ref{timeint}. Note that in \eqref{intHe0} we performed a change of variable. Calculations show that the Hamiltonian $\hat H_{\mathcal{E}_0}^{(\textrm{I})}(t)$ involves interacting gravitons and photons whose wave vectors are aligned in the positive direction of the $z$ axis. This allowed us to express integrals in \eqref{intHe0} using the frequencies of the interacting bosons as variables. 
Substituting \eqref{Hkappaint} and \eqref{intHe0} into \eqref{UE} we finally obtain the time evolution operators $\hat{U}_{\mathcal{E}_0}(t)$ and we can therefore proceed to evaluate the dynamics of the state of the system.
\subsubsection{Time evolution of the state}
We assume that the initial photon-graviton state is the vacuum state, $\lvert\Psi(0)\rangle=\lvert 0\rangle\equiv\lvert 0_{\lambda\bold k}0_{\gamma\bold p}\rangle$. Once the time-evolution starts, at time $t$ the state of the system is found in
\begin{align}
\lvert\Psi(t)\rangle=&\hat{U}_\textrm{I}(t)\lvert\Psi(0)\rangle\nonumber\\
=&\hat{U}_{\mathcal{E}_0^2}(t)\left\{\hat{\mathbb{I}}-\frac{i}{\hbar}\int dt'\left[\hat H_{\mathcal{E}_0}^{(I)}(t')+\hat H_{\kappa}(t')\right]\right\}\lvert 0 \rangle\nonumber\\
=&\hat{U}_{\mathcal{E}_0^2}(t)\bigg[\lvert 0_{\lambda\bold k} 0_{\gamma\bold p} \rangle\nonumber\\
&\left.+\frac{2\pi^2i\,C}{\hbar\,\Omega}\left(\lvert\psi\rangle+\pi\sqrt{\Omega}\,\alpha_0\,t^2\lvert 1_{\uparrow\bm\kappa} 0_{\gamma\bold p} \rangle\right)\right]
\label{psif}
\end{align}
where
\begin{align}
\lvert\psi\rangle=&\frac{\sqrt{2}}{c^3}\int_0^\Omega d\omega \,\omega\left(\lvert1_{\uparrow\omega} 1_{+(\Omega-\omega)}\rangle+\lvert1_{\downarrow\omega} 1_{\times(\Omega-\omega)}\rangle\right)
\label{psi}
\end{align}
and where we made use of the time integrals of the Hamiltonians in \eqref{Hkappaint} and \eqref{intHe0}. 

The final state at the lowest perturbative order therefore encompasses both the displacement of the graviton mode $\omega_{\bold p}=2\Omega$ due to the action of the unitary operator $\hat{U}_{\mathcal{E}_0^2}(t)$, and the superposition of three possible outcomes: the vacuum state, the presence of a single photon with the same polarization and wave vector of the classical beam (backreation of the quantum EM field), and the state $\lvert\psi\rangle$ in \eqref{psi}.
The latter describes a photon-graviton entangled state in the polarization degrees of freedom. 

To present the polarization entanglement in a more evident form, we isolate this outcome. Experimentally, this can be accomplished by means of a pass-band filter. In particular, we set up our tool such that we filter the frequency of the EM wave out and let pass only those photons whose frequency is lower than $\Omega$.
At this point, we normalize the resulting state and rewrite it in the basis of the polarization degrees of freedom:
\begin{align}
\lvert\psi\rangle=\frac{1}{\sqrt{2}}\left(\lvert\uparrow\rangle\lvert+\rangle+\lvert\downarrow\rangle\lvert\times\rangle\right)=\lvert\Psi^{(+)}\rangle,
\end{align}
where in the second equivalence we expressed the state in the Bell state basis \cite{PhysRevLett.87.277902}:
\begin{subequations}
\begin{align}
\lvert\Psi^{(+)}\rangle=\frac{1}{\sqrt{2}}\left(\lvert\uparrow\rangle\lvert+\rangle+\lvert\downarrow\rangle\lvert\times\rangle\right),\\
\lvert\Psi^{(-)}\rangle=\frac{1}{\sqrt{2}}\left(\lvert\uparrow\rangle\lvert+\rangle-\lvert\downarrow\rangle\lvert\times\rangle\right),\\
\lvert\Phi^{(+)}\rangle=\frac{1}{\sqrt{2}}\left(\lvert\uparrow\rangle\lvert\times\rangle+\lvert\downarrow\rangle\lvert+\rangle\right),\\
\lvert\Phi^{(-)}\rangle=\frac{1}{\sqrt{2}}\left(\lvert\uparrow\rangle\lvert\times\rangle-\lvert\downarrow\rangle\lvert+\rangle\right).
\end{align}\label{Bell}
\end{subequations}
In a hypothetical experiment carried out with state-of-the-art technology, we would not be able to access the graviton subsystem. What we could however reveal is the photon state, which is described by the reduced density matrix
\begin{align}
\rho_a\equiv \textrm{Tr}_b[\lvert\psi\rangle\langle\psi\rvert]=\frac{1}{2}(\lvert\uparrow\rangle\langle\uparrow\rvert+\lvert\downarrow\rangle\langle\downarrow\rvert).
\label{rhoa1}
\end{align}
As output of our measurement we would therefore observe single photons prepared in a maximally mixed state in the polarization basis, and whose frequency is expected in the range $0<\omega_\textrm{k}<\Omega$. Such photons are clearly unexpected in a classical theory of gravity, as their presence heralds the entanglement with hidden quantum degrees of freedom stemming from the graviton field. The detection of such heralding photons could therefore represent a signature of the quantum nature of the gravitational field.

Before concluding this section, we further characterize the heralding photons by investigating their spectrum.
In the frequency range $0<\omega_{\bold k}<\Omega$ the spectrum of the expected photons is
\begin{align}
S(\omega)&\equiv\int d\varsigma\langle 0\rvert \hat U^\dag(t)\hat a_{\lambda\bold k}^\dag\hat a_{\lambda\bold k}\hat U^\dag(t)\lvert0\rangle\nonumber\\
&=\int d\varsigma\langle 0\rvert \hat U_{\mathcal{E}_0}^\dag(t)\hat a_{\lambda\bold k}^\dag\hat a_{\lambda\bold k}\hat U_{\mathcal{E}_0}^\dag(t)\lvert0\rangle\nonumber\\
&=\frac{4G(2\pi)^{6}\mathcal{P}\,d\,t}{ c^6\kappa} \omega^2
\end{align}
where $d\varsigma=d\vartheta d\varphi\sin\vartheta$ is the differential solid angle of the emitted photon and $\mathcal{P}=\pi\sigma^2\mathcal{E}_0^2c$ is the power carried by the classical EM wave. This result suggests that pair production phenomena facilitate the generation of high-frequency heralding photons.  

\subsection{Circularly polarized EM wave}
As a second scenario, we consider an EM wave circularly polarized in the $x-y$ plane. This is mathematically described in terms of two classical plane waves dephased by $\pi/2$:
\begin{subequations}\label{classfie2}
\begin{align}
\bm{\mathcal{E}}(\bold x,t)=&\mathcal{E}_0\xi(x,y)\left[\bold e_x \sin(\Omega \,t-\kappa\,z)+\bold e_y\cos(\Omega \,t-\kappa\,z)\right],\\
\bm{\mathcal{B}}(\bold x,t)=&\mathcal{E}_0\xi(x,y)\left[\bold e_x\cos(\Omega \,t-\kappa\,z)-\bold e_y \sin(\Omega \,t-\kappa\,z)\right].
\end{align}
\end{subequations}

We make use of the same procedure as for the linearly polarized EM wave to calculate both the interaction Hamiltonian and the time evolution operator. Once we substitute \eqref{classfie2} and \eqref{quantfie} into \eqref{Ttensor} we obtain the density interaction Hamiltonians
\begin{align}
\mathcal{H}_{\mathcal{E}_0^2}=&\frac{1}{2}\left(\hat h_{22}-\hat h_{11}\right)\left(\mathcal{E}_y^2-\mathcal{E}_x^2\right),\\
\mathcal{H}_{\mathcal{E}_0}=&\left(\hat h_{22}+\hat h_{33}\right)\left(\hat E_x\mathcal{E}_x-c^2\hat B_x\mathcal{B}_x\right)\nonumber\\
&+\left(\hat h_{11}+\hat h_{33}\right)\left(\hat E_y\mathcal{E}_y-c^2\hat B_y\mathcal{B}_y\right)\nonumber\\
&-\hat h_{23}\left(\hat E_z\mathcal{E}_y-c^2\hat B_z\mathcal{B}_y\right)-\hat h_{13}\left(\hat E_z\mathcal{E}_x-c^2\hat B_z\mathcal{B}_x\right)\nonumber\\
&-\hat h_{12}\left(\hat E_y\mathcal{E}_x-c^2\hat B_y\mathcal{B}_x+\hat E_x\mathcal{E}_y-c^2\hat B_x\mathcal{B}_y\right).
\end{align}
Spatial integrals required to obtain the Hamiltonian from the Hamiltonian density are equivalent to those performed in the previous section, therefore we will not report details of such calculations.
Once integrated $\mathcal{H}_{\mathcal{E}_0^2}$ and $\mathcal{H}_{\mathcal{E}_0}$ spatially, and carried out the RWA, we obtain the two Hamiltonian operators
\begin{align}
\hat H_{\mathcal{E}_0^2}^{(\textrm{I})}=&\frac{\mathcal{E}_0^2 \mathcal{A}\pi^2}{2\sqrt{\pi\Omega}}\left(\hat b_{+2\kappa}+\hat b_{+2\kappa}^\dag\right)
\label{HE2lim2}
\end{align}
and
\begin{align}
\hat H_{\mathcal{E}_0}^{(\textrm{I})}(t)=&\frac{\sqrt\hbar\mathcal{E}_0\mathcal{A}V}{(2\pi)^3}\sum_{\lambda,\gamma}\int d^3\bold p\,d^3\bold k\sqrt{\frac{\omega_{\bold k}}{\omega_{\bold p}}}\,\left(\mathcal{I}_{\lambda\bold k}^{\gamma\bold p}+i \mathcal{J}_{\lambda\bold k}^{\gamma\bold p}\right) \nonumber\\
& e^{-i\left(\Omega+\omega_{\bold k}\right)t}\left\{\hat a_{\lambda\bold k}\chi(\bold k_\perp+\bold p_\perp,k_z+p_z)e^{-i\omega_{\bold k}t}\right.\nonumber\\
&\left.-\hat a_{\lambda\bold k}^\dag \chi(\bold k_\perp+\bold p_\perp,-k_z+p_z)e^{i\omega_{\bold k}t}\right\}\hat b_{\gamma\bold p}+\textrm{h.c.}
\label{He0circ}
\end{align}
where
\begin{align}
\mathcal{J}_{\lambda\bold k}^{\gamma\bold p}=&
\left(\epsilon^{\gamma\bold p}_{xx}+\epsilon^{\gamma\bold p}_{zz} \right)e^{\lambda\bold k}_y-\epsilon^{\gamma\bold p}_{xy} e^{\lambda\bold k}_x- \epsilon^{\gamma\bold p}_{yz} e^{\lambda\bold k}_z\nonumber\\
&+\epsilon^{\gamma\bold p}_{xy} (\bar k_ze^{\lambda\bold k}_x-\bar k_xe^{\lambda\bold k}_x)+ \epsilon^{\gamma\bold p}_{xz} (\bar k_xe^{\lambda\bold k}_y-\bar k_ye^{\lambda\bold k}_x)\nonumber\\
&-\left(\epsilon^{\gamma\bold p}_{yy} +\epsilon^{\gamma\bold p}_{zz} \right)(\bar k_x e^{\lambda\bold k}_y-\bar k_y e^{\lambda\bold k}_x).
\label{He0circ}
\end{align}

It is relevant to notice that the Hamiltonian in \eqref{HE2lim2} is identical to the one in \eqref{HE2lim} apart from a factor 2, whereas 
\eqref{He0circ} differs from \eqref{He0} by the presence of $\left(\mathcal{I}_{\lambda\bold k}^{\gamma\bold p}+i \mathcal{J}_{\lambda\bold k}^{\gamma\bold p}\right)$. In the limits we are considering, namely $\kappa\sigma\gg1$ and $\kappa d\gg1$, we again have that the wave vectors $\bold p$ and $\bold k$ are aligned with the $z$ axis, which allows us to simplify the function $\mathcal{J}_{\lambda\bold k}^{\gamma\bold p}$ to
\begin{align}
\mathcal{J}_{\uparrow\bold k}^{+\bold p}=\mathcal{J}_{\downarrow\bold k}^{\times\bold p}=0,\;\;\;\;\textrm{and}\;\;\;\;\mathcal{J}_{\uparrow\bold k}^{\times\bold p}=-\mathcal{J}_{\downarrow\bold k}^{+\bold p}=-\sqrt{2}.
\end{align}

The computation of the time evolution operators \eqref{UE} results therefore immediate:
\begin{widetext}
\begin{subequations}\label{Ucirc}
\begin{align}
\hat{U}_{\mathcal{E}_0^2}(t)=&\hat D_{+2\kappa}[2\alpha(t)],\\
\hat{U}_{\mathcal{E}_0}(t)=&\hat{\mathbb{I}}+\frac{4\pi^3\sqrt{2}\,i\,C\,t^2}{\hbar}\left[(1+i)\hat a_{\uparrow\bm\kappa}^\dag \alpha_0+(1-i)\hat a_{\uparrow\bm\kappa}\alpha^*_0\right]\nonumber\\
&-\frac{(2\pi)^2 \,i\,C}{\hbar\,\Omega\,c^3}\sqrt{2}\left\{\int_0^\infty d\omega \,\sqrt{\omega^3}\sqrt{\omega+\Omega}\left[\hat a_{\uparrow\omega}^\dag \hat b_{+(\Omega+\omega)}+\hat a_{\downarrow\omega}^\dag \hat b_{\times(\Omega+\omega)}+i\left(\hat a_{\uparrow\omega}^\dag \hat b_{\times(\Omega+\omega)}-\hat a_{\downarrow\omega}^\dag \hat b_{+(\Omega+\omega)}\right)\right]\right.\nonumber\\
&\left.-\int_0^\Omega d\omega\,\sqrt{\omega^3}\sqrt{\Omega-\omega}\left[\hat a_{\uparrow\omega}\hat b_{+(\Omega-\omega)}+\hat a_{\downarrow\omega}\hat b_{\times(\Omega-\omega)}+i\left(\hat a_{\uparrow\omega}\hat b_{\times(\Omega-\omega)}-\hat a_{\downarrow\omega}\hat b_{+(\Omega-\omega)}\right)\right]\right\}+\textrm{h.c.}
\end{align}
\end{subequations}
\end{widetext}
The dynamics of the system is therefore expressed via the application of time evolution operator \eqref{time:evolution:operator} and \eqref{Ucirc} to the initial (vacuum) state.
\begin{align}
\lvert\Psi(t)\rangle=&\hat{U}_\textrm{I}(t)\lvert\Psi(0)\rangle\nonumber\\
=&\hat{U}_{\mathcal{E}_0^2}(t)\lvert 0_{\lambda\bold k} 0_{\gamma\bold p} \rangle\nonumber\\
&+\frac{2\pi^2i\,C}{\hbar}\hat{U}_{\mathcal{E}_0^2}(t)\left[\lvert\phi\rangle+\frac{2\pi(1+i)\,\alpha_0\,t^2}{\sqrt{\Omega}}\lvert 1_{\uparrow\bm\kappa} 0_{\gamma\bold p} \rangle\right]
\label{psif2}
\end{align}
where
\begin{align}
\lvert\phi\rangle=&\frac{\sqrt{2}}{\Omega\,c^3}\int_0^\Omega d\omega \,\omega\left[\lvert1_{\uparrow\omega} 1_{+(\Omega-\omega)}\rangle+\lvert1_{\downarrow\omega} 1_{\times(\Omega-\omega)}\rangle\right.\nonumber\\
&\left.+i\left(\lvert1_{\uparrow\omega} 1_{\times(\Omega-\omega)}\rangle-\lvert1_{\downarrow\omega} 1_{+(\Omega-\omega)}\rangle\right)\right]
\label{psi2}
\end{align}

Once we filter the classical beam out by means of the bass-band filter and normalize the state, we obtain
\begin{align}
\lvert\phi\rangle=&\frac{1}{2}\left[\lvert\uparrow\rangle\lvert+\rangle+\lvert\downarrow\rangle\lvert\times\rangle+i\left(\lvert\uparrow\rangle\lvert\times\rangle-\lvert\downarrow\rangle\lvert+\rangle\right)\right]\nonumber\\
=&\frac{1}{\sqrt{2}}\left(\lvert\Psi^{(+)}\rangle+i\lvert\Phi^{(-)}\rangle\right),
\end{align}
where in the second line we again expressed the state in the Bell state basis exposed in \eqref{Bell}.
The photon reduced density matrix is
\begin{align}
\rho_a=\frac{1}{2}(\lvert\uparrow\rangle\langle\uparrow\rvert+\lvert\downarrow\rangle\langle\downarrow\rvert+i\lvert\uparrow\rangle\langle\downarrow\rvert-i\lvert\downarrow\rangle\langle\uparrow\rvert).
\end{align}
Therefore, in contrast to what we have seen in \eqref{rhoa1}, if the polarization of the classical EM wave is circular, the reduce density matrix of the heralding photons displays quantum coherence. In particular, the single photon state can be rearranged in a balanced superposition of the vertical and horizontal polarizations:
\begin{align}
\lvert\psi\rangle_a=\frac{e^{i\varphi}}{\sqrt{2}}\left(\lvert\uparrow\rangle-i\lvert\downarrow\rangle\right),
\end{align}
where $\varphi$ is a global phase.

\section{Can we observe heralding photons?}
As discussed below \eqref{psif}, the output state consists in the superposition of three outcomes: the vacuum state, the entangled state $\lvert\psi\rangle$, and the state $\lvert 1_{\uparrow\bm\kappa} 0_{\gamma\bold p} \rangle$. In the perturbative regime, the vacuum state is expected to be the dominant outcome. The latter describes the back-reaction of the graviton field on the quantum electromagnetic field via the operator $\hat U_{\mathcal{E}_0^2}$. Finally, $\lvert\psi\rangle$ encodes the creation of the photon-graviton pair that are entangled in the polarization degrees of freedom. Although the presence of an opportunely calibrated pass-band filters would facilitate the discrimination of the photon state in \eqref{rhoa1} and, ipso facto, of the state $\lvert\psi\rangle$, the detection of single photons with frequency $\omega_{\bold k}<\Omega$ might result challenging.
For this reason, in this section we want to quantitatively estimate the transition probability for the entangled state described by $\lvert\psi\rangle$ in \eqref{psi} to occur. 

\subsection{Transition probability}
Estimating the probability to obtain $\lvert\psi\rangle$ from the initial state requires to calculating the transition probability to obtain $\lvert\psi\rangle$ at time $t$ from the vacuum, $c_\psi=\langle\psi\rvert\hat{U}_\textrm{I}(t)\vert 0_{\lambda\bold k} 0_{\gamma\bold p} \rangle\equiv\langle\psi\vert\Psi(t)\rangle$. We first notice that the three state $\lvert 0_{\lambda\bold k} 0_{\gamma\bold p} \rangle$, $\lvert 1_{\uparrow\bm\kappa} 0_{\gamma\bold p} \rangle$ and $\lvert\psi\rangle$ in \eqref{psif} are orthogonal. Whereas the orthogonality with the vacuum state is trivial, the orthogonality between $\lvert 1_{\uparrow\bm\kappa} 0_{\gamma\bold p} \rangle$ and $\lvert\psi\rangle$ is guaranteed by the fact that the frequency of both the graviton and the photon in \eqref{psi} is lower than $\Omega$. This suggests that the transition probability we are required to calculate reduces to $\langle\psi\vert\psi\rangle$. 

A convenient way to carry out the transition probability without incurring divergence factors consists in loosening the approximations $\kappa\sigma\gg1$ and $\kappa d\gg1$, and calculating 
\begin{align}
\langle\psi\vert\psi\rangle= \frac{1}{\hbar^2}\int dt'\,dt''\langle\hat H_{\mathcal{E}_0}(t'') \hat H_{\mathcal{E}_0}(t')\rangle
\label{psipsi}
\end{align}
by initially employing the Hamiltonian operator $\hat H_{\mathcal{E}_0}(t)$ given in \eqref{He0}. 
Details of these calculations are illustrated in Appendix \ref{normalization}, whereas here we report the main result:
\begin{align}
c_\psi=&\frac{4(2\pi)^6G \,\mathcal{P}d\,\Omega^2 t}{3c^6}.
\label{cphi}
\end{align}
\subsection{Discussion}
The presence of the factor $4(2\pi)^6G/(3c^6)\sim 10^{-56}$ in \eqref{cphi} distinctly enlightens us about the difficulties of observing photon-graviton entanglement events in an artificial, controllable scenario. Indeed,
the specifics of the electromagnetic wave (which in a controllable environment could be generated by a laser) must fulfill prerequisites in terms of power, frequency, propagation distance and operation time that are hardly achievable by means of state-of-the-art technology. 

To make a concrete example, we consider the parameters employed in the LIGO project for the detection of gravitational waves \cite{aasi2015advanced}. The experimental set-up consists of a Michelson interferometry scheme, whose design has been modified in order to have a Fabry-Perot cavity at each arm of the interferometer. The presence of the Fabry-Perot cavity not only amplifies the power of the confined field to about 750 kW, but it also increases the effective distance traveled by the EM wave dramatically. Indeed, the high finesse ensures around 300 round-trips before the laser beam leaves the cavity, thereby increasing the effective travelled path from about 4 km (real length of the Fabry-Perot cavity) to about 1200 km. Finally, the interferometer is seeded with a Nd:YAG laser at 1064 nm.

Despite the extraordinary features of such experimental set-up, by making use of the parameters listed above and assuming an operation time of about six months, the expected transition probability to obtain the desired entangled state is $c_\psi\sim 10^{-7}$. To enhance this value drastically we would require pump laser with shorter wavelength, as the transition probability depends on the frequency of the electromagnetic wave quadratically. As an example, recent experiments have shown the possibility to generate 4W high-stability CW laser at 355 nm \cite{Wei:24}. Implemented in a LIGO-like set-up, such laser would increase the transition probability to generate photon-graviton entanglement by one order of magnitude. Further improvements might occur by considering larger Fabry-Perot cavities and utilizing mirrors with higher finesse, as this would increase the number of round-trips, and ipso-facto, the effective path length traveled by the EM wave, as well as the power confined in the Fabry-Perot cavity. 

Even though the detection of heralding photons in controllable environments has been proved to be a challenging task, the observation of entanglement events occurring naturally in cosmological phenomena, such as in gamma ray bursts (GRBs) \cite{RevModPhys.76.1143} or pulsar emissions \cite{gedalin2002mechanism}, might result more accessible experimentally. As already pointed out in \cite{lageyreGravitationalInfluenceHigh2022}, the extremely powerful light beams released in GRB events can bring about a non-negligible deformation of the spacetime. However, if on the one hand studying the deformation of the spacetime seems to be a nontrivial operation, detailed investigations of both the spectra and the polarization of the detected radiation might on the other hand reveal the presence of heralding photons, thereby confirming the presence of photon-graviton polarization entanglement. 

\section{Conclusions}

In this work we have investigated the particle production of photon-graviton pairs induced by the propagation of a classical EM wave. We focused on the quantum state of the gravitational and electromagnetic field, showing that photons and gravitons get entangled in the polarization degrees of freedom. We have also discussed the possibility to individuate entangled photons. Since we cannot access the quantum state of gravitons, such photons herald the entanglement with hidden quantum degrees of freedom, thereby revealing the signature of the quantum nature of gravity.

In future works we will further characterize the entanglement between quantum gravitational and electromagnetic modes, for example by taking into account the spectral/temporal features of the classical electromagnetic signal. 
Similarly to what we already observe in controllable environments such as quantum optical set-ups \cite{PhysRevLett.92.127903}, we expect that the spectral profile of the traveling wave could affect the quantum state of the generated particles and induce a form of spectral entanglement.

Finally, we remark that this study offers further perspectives on the role of gravitons in future quantum technologies. 
If the detection of gravitational waves has both further confirmed the robustness of general relativity and contributed to the rise of the multi-messenger astronomy, we expect that the detection of gravitons will revolutionize all aspects of quantum physics, from particle physics to quantum information and computing. In due course, the controllable generation of photon-graviton Bell states will pave the way to quantum computing protocols and communication systems exploiting both electromagnetic and gravitational information carriers. 

\section{Acknowledgment}
The author thanks David Edward Bruschi for valuable comments and feedback, and Paul Lagayre for the interesting discussions.
The author acknowledges support from the joint project No. 13N15685 ``German Quantum Computer based on Superconducting Qubits (GeQCoS)'' sponsored by the German Federal Ministry of Education and Research (BMBF) under the \href{https://www.quantentechnologien.de/fileadmin/public/Redaktion/Dokumente/PDF/Publikationen/Federal-Government-Framework-Programme-Quantum-technologies-2018-bf-C1.pdf}{framework programme
``Quantum technologies -- from basic research to the market''}.

\bibliography{Alexandria}
\appendix
\onecolumngrid
\section{Hamiltonian and time evolution operator}\label{Hamandevo}

The calculation of the interaction Hamiltonian follows the standard procedure. We recall that the interaction Lagrangian density depends on the conjugate momentum of the electromagnetic field. At the lowest order in $h$, the conjugate momentum reads
\begin{align}
\Pi^{\alpha}\equiv\frac{\partial \mathcal{L}}{\partial(\partial_0A_\alpha)}=\Pi_{(0)}^{\alpha}+\Pi_{(1)}^{\alpha},
\end{align}
where $\Pi_{(0)}^{\alpha}$ is the conjugate momentum without interactions, whereas
\begin{align}
\Pi_{(1)}^{\alpha}=\frac{h}{2}F^{\alpha0}+h^{\mu0}\eta^{\alpha\lambda}F_{\mu\lambda}-h^{\mu\alpha}\eta^{0\lambda}F_{\mu\lambda}
\label{pi1}
\end{align}
is the first order conjugate momentum.
The interaction Hamiltonian density therefore corresponds to
\begin{align}
\mathcal{H}_\textrm{I}&=  \Pi_{(1)}^{\alpha}F_{0\alpha}-\mathcal{L}_{(1)}.
\label{HI}
\end{align}
 
Once calculated the interaction Hamiltonian density via \eqref{HI}, we separate the term proportional to $\mathcal{E}_0^2$ from the term proportional to $\mathcal{E}_0$ as follows
\begin{align}
\mathcal{H_\textrm{I}}&=  \mathcal{H}_{\mathcal{E}_0^2}+\mathcal{H}_{\mathcal{E}_0}.
\label{hamden}
\end{align}
The explicit form of each term depends on the polarization of the EM wave. Note that we have omitted all contributions that are not proportional to $\mathcal{E}_0$ or $\mathcal{E}_0^2$. 
The Hamiltonian in the interaction picture is $\hat H_\textrm{I}^{(\textrm{I})}(t)=\hat H_{\mathcal{E}_0^2}^{(\textrm{I})}(t)+\hat H_{\mathcal{E}_0}^{(\textrm{I})}(t)$ where
\begin{align}
\hat H_{\mathcal{E}_0^2}^{(\textrm{I})}(t)=&\int d^3\bold x\,
\mathcal{H}_{\mathcal{E}_0^2}^{(\textrm{I})}(\bold x,t)\\
\hat H_{\mathcal{E}_0}^{(\textrm{I})}(t)=&\int d^3\bold x\,
\mathcal{H}_{\mathcal{E}_0}^{(\textrm{I})}(\bold x,t).
\end{align}

To obtain an explicit form for $\hat{U}_{\mathcal{E}_0}(t)$, we first need to compute $\hat{U}_{\mathcal{E}_0^2}^\dag(t)\hat H_{\mathcal{E}_0}^{(\textrm{I})}\hat{U}_{\mathcal{E}_0^2}(t)$. This leads to
\begin{align}
\hat{U}_{\mathcal{E}_0^2}^\dag(t)\hat H_{\mathcal{E}_0}^{(\textrm{I})}\hat{U}_{\mathcal{E}_0^2}(t)=&H_{\mathcal{E}_0}^{(\textrm{I})}(t)+\hat H_{\kappa}(t),
\end{align}

\section{Time integrals of $\hat H_{\kappa}(t)$ and $\hat H_{\mathcal{E}_0}^{(\textrm{I})}(t)$}\label{timeint}
 To obtain the explicit form of the time evolution operator in \eqref{UE}, we need to calculate the time integrals of both $\hat H_{\kappa}(t)$ and $\hat H_{\mathcal{E}_0}^{(\textrm{I})}(t)$. Starting from the first Hamiltonian, we have
\begin{align}
\int_0^t dt' \hat H_{\kappa}(t')=&-\pi^3\sqrt{2\hbar}\mathcal{E}_0\mathcal{A} V\int dt'\left[\hat a_{\uparrow\Omega}^\dag\alpha(t')+\hat a_{\uparrow\Omega}\alpha^*(t')\right]-\frac{\pi^3\sqrt{2\hbar}\mathcal{E}_0\mathcal{A} V t}{2}\left[\hat a_{\uparrow\Omega}^\dag\alpha(t)+\hat a_{\uparrow\Omega}\alpha^*(t)\right]=\frac{t}{2}\hat H_{\kappa}(t).
\end{align}

We now focus on
\begin{align}
\int_0^t dt' \hat H_{\mathcal{E}_0}^{(\textrm{I})}(t')\simeq&2\pi C\sum_{\lambda,\gamma}\int d^3\bold p\,d^3\bold k\sqrt{\frac{\omega_{\bold k}}{\omega_{\bold p}}}\,\mathcal{I}_{\lambda\bold k}^{\gamma\bold p}e^{-i\left(\kappa-p_z \right)\frac{d}{2}}\delta(k_x+p_x)\delta(k_y+p_y)\nonumber\\
& \left\{\hat a_{\lambda\bold k}e^{\frac{ik_zd}{2}}\delta(\kappa-k_z-p_z)\delta\left(\Omega-\omega_{\bold k}-\omega_{\bold p}\right)-\hat a_{\lambda\bold k}^\dag e^{-\frac{ik_zd}{2}}\delta(\kappa+k_z-p_z)\delta\left(\Omega+\omega_{\bold k}-\omega_{\bold p}\right)\right\}\hat b_{\gamma\bold p}+\textrm{h.c.}
\end{align}
where the delta functions over frequencies emerge by taking the limits $t\gg (\Omega-\omega_{\bold k}-\omega_{\bold p})^{-1}$ and $t\gg (\Omega+\omega_{\bold k}-\omega_{\bold p})^{-1}$.
We want to rewrite this expression in a more convenient way. We start by calculating the integrals over $p_x$ and $p_y$:

\begin{align}
\int_0^t dt' \hat H_{\mathcal{E}_0}^{(\textrm{I})}(t')=&\frac{2\pi C}{c}\sum_{\lambda,\gamma} \int dp_z\,d^3\bold k\,\frac{\sqrt[4]{k_x^2+k_y^2+k_z^2}}{\sqrt[4]{k_x^2+k_y^2+p_z^2}}\,\mathcal{I}_{\lambda\bold k}^{\gamma\bold p}e^{-i\left(\kappa-p_z \right)\frac{d}{2}}\nonumber\\
& \left\{\hat a_{\lambda\bold k}e^{\frac{ik_zd}{2}}\delta(\kappa-k_z-p_z)\delta\left(\kappa-\sqrt{k_x^2+k_y^2+k_z^2}-\sqrt{k_x^2+k_y^2+p_z^2}\right)\right.\nonumber\\
&\left.-\hat a_{\lambda\bold k}^\dag e^{-\frac{ik_zd}{2}}\delta(\kappa+k_z-p_z)\delta\left(\kappa+\sqrt{k_x^2+k_y^2+k_z^2}-\sqrt{k_x^2+k_y^2+p_z^2}\right)\right\}\hat b_{\gamma\bold p}+\textrm{h.c.}
\end{align}
The integral over $p_z$ leads to
\begin{align}
\int_0^t dt' \hat H_{\mathcal{E}_0}^{(\textrm{I})}(t')=&\frac{2\pi C}{c}\sum_{\lambda,\gamma} \int d^3\bold k\left\{\hat a_{\lambda\bold k}\hat b_{\gamma\bold p'}\frac{\sqrt[4]{k_x^2+k_y^2+k_z^2}}{\sqrt[4]{k_x^2+k_y^2+(\kappa-k_z)^2}}\,\mathcal{I}_{\lambda\bold k}^{\gamma\bold p'}\delta\left(\kappa-\sqrt{k_x^2+k_y^2+k_z^2}-\sqrt{k_x^2+k_y^2+(\kappa-k_z)^2}\right)\right.\nonumber\\
&\left.-\hat a_{\lambda\bold k}^\dag \hat b_{\gamma\bold p''}\frac{\sqrt[4]{k_x^2+k_y^2+k_z^2}}{\sqrt[4]{k_x^2+k_y^2+(\kappa+k_z)^2}}\,\mathcal{I}_{\lambda\bold k}^{\gamma\bold p''}\delta\left(\kappa+\sqrt{k_x^2+k_y^2+k_z^2}-\sqrt{k_x^2+k_y^2+(\kappa+k_z)^2}\right)\right\}+\textrm{h.c.}
\end{align}
where we have defined the graviton wave vectors as $\bold p'=(-k_x,-k_y,\kappa-k_z)$ and $\bold p''=(-k_x,-k_y,\kappa+k_z)$. Conveniently, we express the integral over $\bold k$ in cylindrical coordinates
\begin{align}
\int_0^t dt' \hat H_{\mathcal{E}_0}^{(\textrm{I})}(t')=&\frac{2\pi C}{c}\sum_{\lambda,\gamma} \int_0^{2\pi} d\vartheta\int dk_z\int_0^\infty d\varrho\,\varrho\left\{\hat a_{\lambda\bold k}\hat b_{\gamma\bold p'}\frac{\sqrt[4]{\varrho^2+k_z^2}}{\sqrt[4]{\varrho^2+(\kappa-k_z)^2}}\,\mathcal{I}_{\lambda\bold k}^{\gamma\bold p'}\delta\left(\kappa-\sqrt{\varrho^2+k_z^2}-\sqrt{\varrho^2+(\kappa-k_z)^2}\right)\right.\nonumber\\
&\left.-\hat a_{\lambda\bold k}^\dag \hat b_{\gamma\bold p''}\frac{\sqrt[4]{\varrho^2+k_z^2}}{\sqrt[4]{\varrho^2+(\kappa+k_z)^2}}\,\mathcal{I}_{\lambda\bold k}^{\gamma\bold p''}\delta\left(\kappa+\sqrt{\varrho^2+k_z^2}-\sqrt{\varrho^2+(\kappa+k_z)^2}\right)\right\}+\textrm{h.c.}
\end{align}
We now notice that the remaining delta functions depend on two variables. By expressing the current variables $\rho$ and $k_z$ in terms of two new variables $w$ and $v$ as $\rho=\sqrt{w^2-v^2}$ and $k_z=v$, we have
\begin{align}
\int_0^t dt' \hat H_{\mathcal{E}_0}^{(\textrm{I})}(t')=&\frac{2\pi C}{c}\sum_{\lambda,\gamma} \int_0^{2\pi} d\vartheta\int_0^\infty dw \int_{\lvert v\rvert\le\lvert w\rvert} dv\,\sqrt{w^3}\left\{\hat a_{\lambda\bold k}\hat b_{\gamma\bold p'}\frac{1}{\sqrt[4]{w^2+\kappa^2-2\kappa v}}\,\mathcal{I}_{\lambda\bold k}^{\gamma\bold p'}\delta\left(\kappa-w-\sqrt{w^2+\kappa^2-2\kappa v}\right)\right.\nonumber\\
&\left.-\hat a_{\lambda\bold k}^\dag \hat b_{\gamma\bold p''}\frac{1}{\sqrt[4]{w^2+\kappa^2+2\kappa v}}\,\mathcal{I}_{\lambda\bold k}^{\gamma\bold p''}\delta\left(\kappa+w-\sqrt{w^2+\kappa^2+2\kappa v}\right)\right\}+\textrm{h.c.}\nonumber\\
=&\frac{2\pi C}{c\kappa}\sum_{\lambda,\gamma} \int_0^{2\pi} d\vartheta\left\{\int_0^\kappa dw \int_{\lvert v\rvert\le\lvert w\rvert} dv\,\sqrt{w^3}\frac{\lvert w-\kappa\rvert}{\sqrt[4]{w^2+\kappa^2-2\kappa v}}\hat a_{\lambda\bold k}\hat b_{\gamma\bold p'}\,\mathcal{I}_{\lambda\bold k}^{\gamma\bold p'}\delta\left(v-w\right)\right.\nonumber\\
&\left.-\int_0^\infty dw \int_{\lvert v\rvert\le\lvert w\rvert} dv\,\sqrt{w^3}\frac{\lvert w+\kappa\rvert}{\sqrt[4]{w^2+\kappa^2+2\kappa v}}\hat a_{\lambda\bold k}^\dag \hat b_{\gamma\bold p''}\,\mathcal{I}_{\lambda\bold k}^{\gamma\bold p''}\delta\left(v-w\right)\right\}+\textrm{h.c.}
\end{align}
where we exploited the composition role of the delta function.
We recall that $v=k_z$ and $w\equiv\sqrt{\rho^2+k_z^2}=\sqrt{k_x^2+k_y^2+k_z^2}\equiv\lvert k\rvert$. The delta function therefore suggests that $\lvert k\rvert=k_z$. In other words, under the limits $\kappa\sigma\gg1$ and $\kappa d\gg1$ processes described by the Hamiltonian $\hat H_{\mathcal{E}_0}^{(\textrm{I})}(t)$ only occurs along the $z$-axis, and the wave vectors of both gravitons and photons are oriented in the same direction of the wave vector of the classical wave. This allows us to drastically simplify the functions $\mathcal{I}_{\lambda\bold k}^{\gamma\bold p'}$ and $\mathcal{I}_{\lambda\bold k}^{\gamma\bold p''}$, as they reduce to
\begin{align}
\mathcal{I}_{\uparrow\bold k}^{+\bold p}=\mathcal{I}_{\downarrow\bold k}^{\times\bold p}=-\sqrt{2},\;\;\;\;\;\textrm{and}\;\;\;\;\;\mathcal{I}_{\uparrow\bold k}^{\times\bold p}=\mathcal{I}_{\downarrow\bold k}^{+\bold p}=0.
\label{condI}
\end{align}
when $\bold k$ and $\bold p$ are aligned in the $z$ direction. Once we perform the integral over $v$ and redefine $w$ in terms of the frequency $\omega$, we have
\begin{align}
\int_0^t dt' \hat H_{\mathcal{E}_0}^{(\textrm{I})}(t')=&\frac{(2\pi)^2 C}{\Omega\,c^3}\sqrt{2}\left\{\int_0^\infty d\omega \,\sqrt{\omega^3}\sqrt{\omega+\Omega}\left[\hat a_{\uparrow\omega}^\dag \hat b_{+(\Omega+\omega)}+\hat a_{\downarrow\omega}^\dag \hat b_{\times(\Omega+\omega)}\right]\right.\nonumber\\
&\left.-\int_0^\Omega d\omega\,\sqrt{\omega^3}\sqrt{\Omega-\omega}\left[\hat a_{\uparrow\omega}\hat b_{+(\Omega-\omega)}+\hat a_{\downarrow\omega}\hat b_{\times(\Omega-\omega)}\right]\right\}+\textrm{h.c.}
\end{align}

\section{Normalization of $\lvert\Psi(t)\rangle$ and calculation of the transition probability amplitude}\label{normalization}

The state we want to normalize is 
\begin{align}
\lvert\Psi(t)\rangle=&\hat{U}_\textrm{I}(t)\lvert\Psi(0)\rangle=\hat{U}_{\mathcal{E}_0^2}(t)\left\{\hat{\mathbb{I}}-\frac{iC}{\hbar}\int_0^{t}dt'\hat{H}_{\textrm{eff}}(t')-\frac{C^2}{\hbar^2}\int_0^t dt'\hat{H}_{\textrm{eff}}(t')\int_0^{t'} dt''\hat{H}_{\textrm{eff}}(t'')\right\}\lvert 0 \rangle\nonumber\\
=&\hat{U}_{\mathcal{E}_0^2}(t)\left[\lvert 0\rangle+\lvert\Psi^{(1)}(t)\rangle+\lvert\Psi^{(2)}(t)\rangle\right].
\end{align}
Up to the second order in $C$, such state results already normalized. Indeed, the normalization of such state reads
\begin{align}
\mathcal{N}\equiv&\langle\Psi(t)\vert\Psi(t)\rangle=\langle0\rvert\hat U_\textrm{I}^\dag(t)\hat U_\textrm{I}(t)\lvert0\rangle=\langle0\rvert\hat{U}_{\mathcal{E}_0}^\dag(t)\hat{U}_{\mathcal{E}_0}(t)\lvert0\rangle=1+\langle\Psi^{(1)}(t)\vert\Psi^{(1)}(t)\rangle+\langle 0\vert\Psi^{(2)}(t)\rangle+\langle\Psi^{(2)}(t)\vert 0\rangle
\nonumber\\
=&1+\frac{C^2}{\hbar^2}\int_0^t dt'\,dt''\langle 0 \rvert\hat H_{\textrm{eff}}(t'') \hat H_{\textrm{eff}}(t')\lvert 0\rangle-\frac{C^2}{\hbar^2}\int_0^t dt'\int_0^{t'} dt''\langle 0 \rvert\left\{\hat{H}_{\textrm{eff}}(t'),\hat{H}_{\textrm{eff}}(t'')\right\}\lvert 0\rangle,
\label{norm}
\end{align}
where $\{\cdot,\cdot\}$ indicates the anticommunator. Once inserted \eqref{Heff2} into the equation above, expressed $\hat H_{\mathcal{E}_0}^{(\textrm{I})}(t)$ and $\hat H_{\kappa}(t)$ via \eqref{He0} and \eqref{Hkappa} respectively, and performed all time integrals, it is straightforward to see that the second and the third terms in \eqref{norm} cancel out. 

To obtain an explicit form of \eqref{psipsi}, we need to simplify the following expression
\begin{align}
(2\pi)^6\sum_{\lambda\gamma}\sum_{\bar\lambda\bar\gamma}\int dt' dt''\int d^3\bold p\,d^3\bold k\, d^3\bar{\bold p}\,d^3\bar{\bold k}\sqrt{\frac{\omega_{\bold k}\omega_{\bar{\bold k}}}{\omega_{\bold p}\omega_{\bar{\bold p}}}}\,\mathcal{I}_{\lambda\bold k}^{\gamma\bold p}\,\mathcal{I}_{\bar\lambda\bar{\bold k}}^{\bar\gamma\bar{\bold p}} e^{-\left[(k_x+ p_x)^2+(k_y+ p_y)^2+(\bar k_x+ \bar p_x)^2+(\bar k_y+ \bar p_y)^2\right]\sigma^2}e^{-i\left(\bar k_z+\bar p_z-k_z-p_z \right)\frac{d}{2}}\nonumber\\
\sinc\left[\left(k_z+p_z-\kappa\right)\frac{d}{2}\right]\sinc\left[\left(\bar k_z+\bar p_z-\kappa\right)\frac{d}{2}\right]\,e^{i\left(\Omega-\omega_{\bold k}-\omega_{\bold p}\right)t'}e^{-i\left(\Omega-\omega_{\bar{\bold k}}-\omega_{\bar{\bold p}}\right)t''}\delta(\bold p-\bar{\bold p})\delta(\bold k-\bar{\bold k})\delta_{\gamma\bar\gamma}\delta_{\lambda\bar\lambda}.
\end{align}
The presence of the last delta functions and Kronecker deltas reduces it drastically, as we obtain
\begin{align}
(2\pi)^6\int dt' dt''\sum_{\lambda\gamma}\int d^3\bold p\,d^3\bold k\, \frac{\omega_{\bold k}}{\omega_{\bold p}}\,\left(\mathcal{I}_{\lambda\bold k}^{\gamma\bold p}\right)^2\,e^{i\left(\Omega-\omega_{\bold k}-\omega_{\bold p}\right)(t'-t'')} e^{-2\left[(k_x+ p_x)^2+(k_y+ p_y)^2\right]\sigma^2}\sinc^2\left[\left(k_z+p_z-\kappa\right)\frac{d}{2}\right].
\end{align}
In the usual limits $\kappa\sigma\gg1$ and $\kappa d\gg1$ we can express both sinc and Gaussian functions in terms of delta function
\begin{align}
(2\pi)^6\frac{\pi^3}{V}\sum_{\lambda\gamma}\int dt' dt''\int d^3\bold p\,d^3\bold k\, \frac{\omega_{\bold k}}{\omega_{\bold p}}\,e^{i\left(\Omega-\omega_{\bold k}-\omega_{\bold p}\right)(t'-t'')}\left(\mathcal{I}_{\lambda\bold k}^{\gamma\bold p}\right)^2\delta(k_x+p_x)\delta(k_y+p_y)\delta(\kappa-k_z-p_z).
\end{align}
Note that, in the limit $t\gg\left(\Omega-\omega_{\bold k}-\omega_{\bold p}\right)^{-1}$, we can also express the integral over $dt'$ in terms of a delta function
\begin{align}
\frac{(2\pi)^{10}}{8V} \sum_{\lambda\gamma}\int dt''\int d^3\bold p\,d^3\bold k\, \frac{\omega_{\bold k}}{\omega_{\bold p}}\,\left(\mathcal{I}_{\lambda\bold k}^{\gamma\bold p}\right)^2\,e^{-i\left(\Omega-\omega_{\bold k}-\omega_{\bold p}\right)t''}
\delta(k_x+p_x)\delta(k_y+p_y)\delta(\kappa-k_z-p_z)\delta\left(\Omega-\omega_{\bold k}-\omega_{\bold p}\right).
\end{align}
We now carry out the integral over $p_x$ and $p_y$ in order to obtain
\begin{align}
&\frac{(2\pi)^{10}}{8cV} \sum_{\lambda\gamma}\int dt''\int dp_z\,d^3\bold k\, \frac{\omega_{\bold k}}{\omega_{\bold p}}\,e^{-i\left(\Omega-\omega_{\bold k}-\omega_{\bold p}\right)t''}\left(\mathcal{I}_{\lambda\bold k}^{\gamma\bold p}\right)^2\delta(\kappa-k_z-p_z)\,\delta\left(\kappa-\sqrt{k_x^2+k_y^2+k_z^2}-\sqrt{k_x^2+k_y^2+p_z^2}\right)
\end{align}
In the last delta function we expressed the dispersion relations for both $\omega_{\bold k}$ and $\omega_{\bold p}$ explicitly in order to show that we carried out the integrals over $p_x$ and $p_y$. 

We rewrite the integrals over $k_x$ and $k_y$ in polar coordinates
\begin{align}
\frac{(2\pi)^{11}}{8c V} \sum_{\lambda\gamma}\int dt''\int dp_z dk_z\delta(\kappa-k_z-p_z)\int_0^\infty d\varrho\varrho\, \frac{\sqrt{\varrho^2+k_z^2}}{\sqrt{\varrho^2+p_z^2}}\left(\mathcal{I}_{\lambda\bold k}^{\gamma\bold p}\right)^2 e^{i\left(\sqrt{\varrho^2+k_z^2}+\sqrt{\varrho^2+p_z^2}-\kappa\right)ct''}\nonumber\\
\delta\left(\kappa-\sqrt{\varrho^2+k_z^2}-\sqrt{\varrho^2+p_z^2}\right),
\end{align}
and perform the integral over $p_z$
\begin{align}
\frac{(2\pi)^{11}}{8c V}\sum_{\lambda\gamma} \int dt''\int dk_z \int_0^\infty d\varrho\varrho\, \frac{\sqrt{\varrho^2+k_z ^2}}{\sqrt{\varrho^2+\left(\kappa-k_z \right)^2}}\left(\mathcal{I}_{\lambda\bold k}^{\gamma\bold p}\right)^2e^{i\left(\sqrt{\varrho^2+k_z^2}+\sqrt{\varrho^2+\left(\kappa-k_z \right)^2}-\kappa\right)ct''}\nonumber\\
\delta\left(\kappa-\sqrt{\varrho^2+k_z^2}-\sqrt{\varrho^2+\left(\kappa-k_z \right)^2}\right).
\end{align}
We introduce two new variables $w$ and $v$ defined in terms of the current variables $\rho$ and $k_z$ as $\rho=\sqrt{w^2-v^2}$ and $k_z=v$, thereby having
\begin{align}
&\frac{(2\pi)^{11}}{8c V} \sum_{\lambda\gamma}\int dt''\int_0^\infty dw \int_{\lvert v\rvert\le\lvert w\rvert} dv  \left(\mathcal{I}_{\lambda\bold k}^{\gamma\bold p}\right)^2\frac{w^2}{\sqrt{w^2+\kappa^2-2\kappa v}}e^{i\left(\sqrt{w^2+\kappa^2-2\kappa v}+w-\kappa\right)ct''}\delta\left(\sqrt{w^2+\kappa^2-2\kappa v}+w-\kappa\right)\nonumber\\
=&\frac{(2\pi)^{11}}{8c\kappa V}  \sum_{\lambda\gamma}\int dt''\int_0^\kappa dw \int_{\lvert v\rvert\le\lvert w\rvert} dv  \left(\mathcal{I}_{\lambda\bold k}^{\gamma\bold p}\right)^2\frac{w^2\lvert \kappa-w\rvert}{\sqrt{w^2+\kappa^2-2\kappa v}}e^{i\left(\sqrt{w^2+\kappa^2-2\kappa v}+w-\kappa\right)ct''}\delta\left(v-w\right)\nonumber\\
=&\frac{(2\pi)^{11}t}{2 c\kappa V}   \int_0^\kappa dw w^2\nonumber\\
=&\frac{(2\pi)^{11} \kappa^2 t}{6c V}  
\end{align}
We can use this result to finally obtain the explicit form of the transition probability in \eqref{psipsi}. We have
\begin{align}
\langle\psi\vert\psi\rangle= \frac{1}{\hbar^2}\int dt'\,dt''\langle\hat H_{\mathcal{E}_0}(t'') \hat H_{\mathcal{E}_0}(t')\rangle=\frac{(2\pi)^5\,\mathcal{E}^2_0V\mathcal{A}^2\kappa^2 t}{6\hbar\, c}=\frac{4(2\pi)^6G \,\mathcal{E}_0^2V \kappa^2 t}{3c^3}=\frac{4(2\pi)^6G \,\mathcal{P}d\,\Omega^2 t}{3c^6}.
\end{align}

\end{document}